# Physicochemical Features and Peculiarities of Interaction of Antimicrobial Peptides with the Membrane

M. Pirtskhalava* (m.pirtskhalava@lifescience.org.ge),
B. Vishnepolsky (b.vishnepolsky@lifescience.org.ge),
M. Grigolava (maia.grigolava@science.org.ge), Grigol Managadze (gmngdz@gmail.com)
I. Beritashvili Center of Experimental Biomedicine, Tbilisi 0160, Georgia
*Corresponding author: m.pirtskhalava@lifescience.org.ge, Tel:: (995) 574 162397

## Abstract

Antimicrobial peptides (AMPs) are anti-infectives that have potential as a novel and untapped class of biotherapeutics. Modes of action of antimicrobial peptides imply interaction with cell envelope (cell wall, outer- and inner-membrane). Comprehensive understanding of peculiarities of interactions of antimicrobial peptides with cell envelope is necessary to perform the task-oriented design of new biotherapeutics, against which for microbes it's hard to work out resistance.

In order to enable a *de novo* design with low costs and in high throughput, *in silico* predictive models have to be required. To develop the performant predictive model, comprehensive knowledge on mechanisms of action of AMPs has to be possessed. The last knowledge will allow us to encode amino acid sequences expressively and to get success to the choosing of the accurate classifier of AMPs.

A shared protective layer of microbial cells is inner, plasmatic membrane. The interaction of AMP with a biological membrane (native and/or artificial) is the most comprehensively studied. We provide a review of mechanisms and results of interaction of AMP with the cell membrane, relying on the survey of physicochemical, aggregative and structural features of AMPs. Potency and mechanism of action of AMP have presented in the terms of amino acid compositions and distributions of the polar and apolar residues along the chain, that is in such physicochemical features of peptides as the hydrophobicity, hydrophilicity, and amphiphilicity. The survey of current data emerges the topics that should be taken into account to get a comprehensive explanation of the mechanisms of action of AMP and to uncover the physicochemical faces of peptides essential to perform their function.

Many different approaches were used to classify AMPs. The survey of the knowledge on sequences, structures, and modes of actions of AMP, allows concluding that, only the physicochemical features of AMPs give the capability to perform the unambiguous classification. Comprehensive knowledge of physicochemical features of AMP is necessary to develop task-oriented methods of design of peptide-based antibiotics *de novo*.

## 1. Introduction

Antimicrobial peptides (AMPs) have potential as a novel class of biotherapeutics. A common feature of AMP is the capability to interact selectively with a microbial envelope. The composition and morphology of cell envelopes vary from the microbe to the microbe. Targets of AMPs for interaction can be many different molecules or molecular complexes of the envelope. Some AMPs even have several targets. Moreover, the type of target changes depending on pH, salt concentration and other environmental factors. Consequently, although AMPs as a membrane-active peptides share two main properties, such as amphipathicity and positive charge, the existence of different targets at the plasma membrane and cell wall explains the wide spectrum of physicochemical features expressed by amino acid sequences of AMPs. Indeed amino acid sequences and 3D structures of AMPs are highly variable. So, AMPs are not designed by nature to interact with a specific target and as a consequence, there is no single mechanism of action. A variety of AMPs and their capability to fulfill their function through many different modes of action is a clue to the closing the easy way to develop the resistance by microbes. Consequently, AMPs became attractive to combat multidrug resistance threat. Although the variety of the physicochemical properties of AMP and their modes of action profitable to deal with the problem of resistance, the same makes it difficult to understand the modes of action to perform an accurate classification and so, to perform *de novo* design of antimicrobials with required properties. In order to enable a *de novo* design with low costs and in high throughput, *in silico* predictive models have to be required. To develop a performant predictive model a comprehensive knowledge of mechanisms of action of AMPs has to be possessed. The last knowledge will allow us to encode amino acid sequences expressively and to get success to the choosing of an accurate classifier. We'll overview the current knowledge on amino acid sequences, secondary structures, physicochemical properties and mode of actions of AMPs to consider the convenience of used classifications.

## 2. Interaction with envelope

Interaction with an envelope, as a rule, is beginning by binding to the outer layer of the envelope which can be followed by either inhibition of the vital pathways in the outer layer of envelopes or passing through it and reaching the plasma membrane. Taking into consideration the morphology of cell wall of gram-positive bacteria and fungi, it's supposed that the majority of AMPs reveal the capability to overcome the outer layer barrier of last organisms and the plasma membrane is the main target for peptides. In the case of gram-negative bacteria, an outer barrier is also a lipid bilayer. So it's understandable that studies of the mechanisms of action of AMPs were mainly concentrated on the exploration of the modes of interactions of AMP with membranes (artificial or natural). Conventionally, when are talking about modes of action are really implied the results of the interaction of AMPs with the membrane which can be the cause of the death of a microbe. Knowledge on AMP- membrane interaction is the most comprehensive.

Relying on the current knowledge, AMPs could be classified into three different categories depending on the results of binding to the membrane. The first category includes AMPs that mainly associated with the hydrocarbon region. The second category of AMPs associate with interface region and interact with polar headgroups and hydrocarbon region of bilayer simultaneously. The third category of AMPs don't associate with the membrane, but pass through it and find their target inside cell.

The capability of the peptide to associate to the particular region or to pass through membrane is determined by their physicochemical features ( expressed by their amino acid sequence) and by the composition of a lipid bilayer. Peptides' concentration is also an important factor that can determine the results of AMP-membrane interaction. The particular peptide may change their behaviour depending on their concentration or bilayer's composition. For instance, cell-penetrating peptides at a certain concentration are acquired antimicrobial potency (Palm, Netzereab and Hallbrink 2006).

So, effective concentration on the membrane surface determines AMPs' ability to cause the perturbation in the three-dimensional structure of lipid bilayer. As a rule, phospholipids composition determines the type of phases and three-dimensionally ordered structures of a bilayer (Epand, Savage and Epand 2007) and so influence the behavior of AMP. The composition and structure of the cell wall can be a determinant of AMP's concentration on the lipid bilayer surface also and consequently, of the mode of interaction with the plasma membrane.

It can be distinguished particular steps at the interaction with the envelope. In the case of cationic AMPs, these steps are: attachment to the cell wall ( gram-negative bacteria, fungi) or outer membrane (gram-negative bacteria) surface by electrostatic interactions; reaching plasmatic membrane and insertion into the lipid bilayer because of hydrophobicity; self-aggregation or forming of the aggregates with lipids in the case of peptides with the propensity to aggregation in the lipid environment; and at the end appearance of either defects (permanent or transient) in the morphology of membrane with accompanying leakage or a reversible (weak) changes and the passing through the lipid bilayer without leakage.

AMP's physicochemical features and the compositions of the cell wall or outer membrane determine AMP's concentration on the plasma membrane surface and mode of interaction with membrane. So, the subtle balance between physicochemical properties of peptides and compositions of the cell wall and lipid bilayer (outer and/or inner membrane) determines the mode of action of AMP. Studies of the relationships between AMP's physicochemical properties (PCP), compositions of the cell wall and/or lipid bilayer and results of interaction with the envelope have uncovered many aspects of the mechanisms of action. In this review, we survey the achievements at the understanding of the peculiarities of interactions of antimicrobial peptides with the lipid bilayers and move out the tasks that should be solved to get a comprehensive explanation of the mechanisms of action that could be applied in the designing of AMPs.

## 3. Physicochemical properties

The main target of AMP at the cell envelope can be considered a plasmatic membrane. Generally, as a major factors that determine the mode of AMP interaction with membrane supposed to be physicochemical properties (PCP) of peptides, that reflect peculiarities of the amino acid composition, distribution of hydrophilic and hydrophobic residues along the chain and 3D structure. Peculiarities of amino acid sequences provide flexibility and structural adaptability of AMPs, that, by turn are responsible for different modes of action on the membrane bilayers. Although other determinants should also be considered, such as the peptide concentration, and the physicochemical properties of the membrane (Bechinger 2015). So, a delicate balance of physical interactions with the membrane is responsible for the mode of action of the peptide. Anyway, knowledge of the physicochemical properties of a peptide is essential to understand modes of interactions with membrane and to predict the results accomplished after interactions.

### 3.1 Amino acids composition and distribution

The amino acid composition can give valuable information concerning to physicochemical features, such as hydrophobicity and charge of peptide. Another valuable feature of AMP is amphipathicity, that requires a knowledge of sequence or even structure to be assessed. It's interesting to know the peculiarities of amino acid compositions and residue distributions in the amino acid sequences of AMPs. To assess the composition and distribution of amino acids in different sets of AMPs, data of the DBAASP database (Pirtskhalava et al. 2016) has been explored. To look for the peculiarities of AMP sequences, it's reasonable to use the data of peptides that were under evolutionary pressure. It means to use ribosomal peptides only.

**Composition.** To reveal functionally valuable peculiarities of the amino acid composition of AMPs, the assessments have to be compared with the assessments of the "average" protein, which is a protein with indefinite function. As a composition of average protein, the amino acid composition of the UniProt database (UniProt Consortium 2019) has been used. The last database is a repository of proteins of many different functions, and so, can be supposed that its amino acid composition corresponds to "average" protein where an impact of evolutionary pressure has smoothed.

The data of 2568 ribosomal peptides has been retrieved from the *DBAASP* database. The differences between amino acid compositions of AMPs' ribosomal set and UniProt is presented in Fig.1. The peculiarity of the amino acid composition of AMPs is an abundance of bulky hydrophobic amino acids ( Phe, Ile, Trp) and also residues such as Cys, Gly, and Lys.

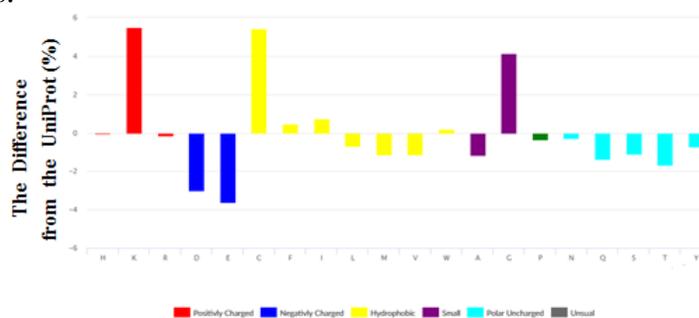

**Fig1.** Amino acid composition of ribosomal peptides of *DBAASP* presenting as the difference from the UniProt (https://dbaasp.org/statistics)

Set of ribosomal AMPs constitute of linear peptides (1443), Cyclic ( N and C termini are covalently linked ) peptides (123) and peptides containing intra-chain covalent bonds. The majority of peptides from the last class are disulfide-bonded peptides (1095). Worth noting, that many cyclic peptides contain disulfide bonds also. According to Fig 2a, 2b and 2c, amino acid compositions of different classes of ribosomal peptides are distinguished from each other. At the same time, shared peculiarities are clearly seen between linear and disulfide-bonded peptides. An abundance of Lys and Gly and a low level of acidic amino acids can be represented as a common property.

It should be emphasized, that an abundance of bulky hydrophobic amino acids ( Phe, Ile, Leu, Trp) and His is an intrinsic feature of linear AMPs, only. We have to note that Phe, Trp and His are aromatic at the same time. Cyclic and disulfide-bonded peptides are rich with Cys, Lys, and Gly. Hallmark of Cyclic peptides is a high percentage of Pro, Ser, and Thr. The last fact can be explained by the requirement of many turns and bends to form the cyclic structure. Worth noting that ribosomal cyclic peptides possess the least total positive charge among AMPs.

It's very interesting the results concerning compositions of two basic amino acids. The portion of Lys in the sequences of ribosomal AMP is higher than it is at the average protein, while the portion of Arg is lower. AMP is mainly a cationic peptide and why their positive charge is majorly provided by the Lys is a question to be answered. We will try to answer on last question below when the differences in the mode of binding of the guanidinium group of Arg and the amino group of Lys with membrane be considered.

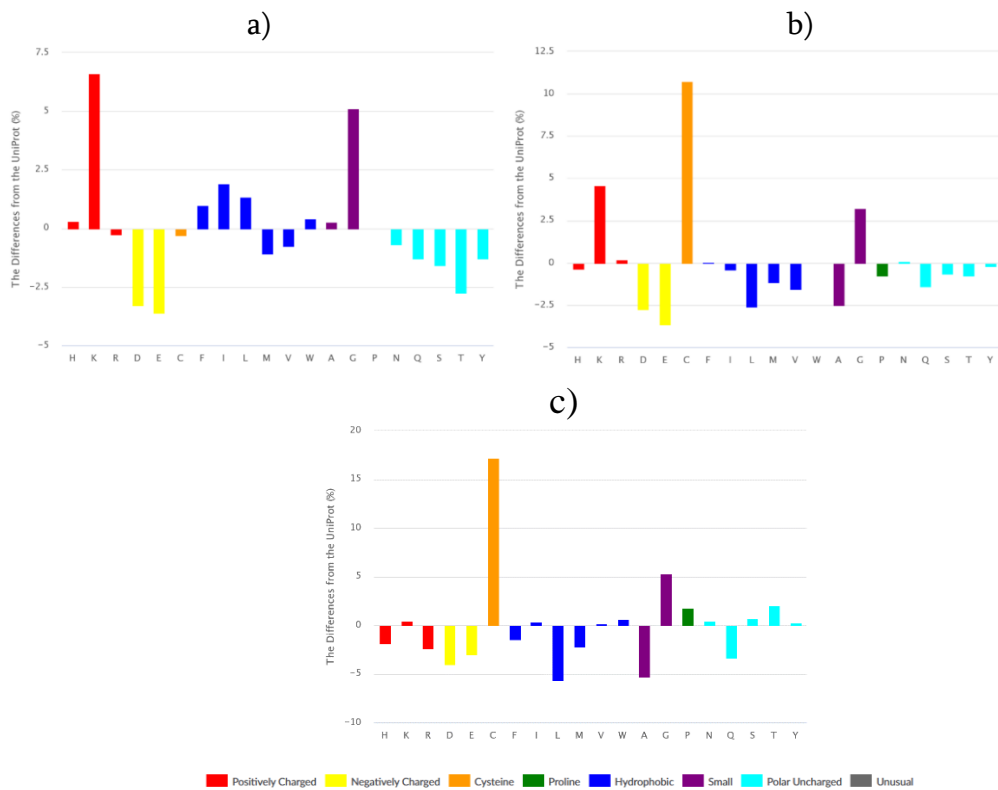

**Fig 2.** Amino acid composition of a) linear; b) disulfide bonded; and c) cyclic ribosomal peptides of *DBAASP* presenting as the difference from the UniProt (https://dbaasp.org/statistics)

**Distribution along the chain.** Certain physicochemical features of peptides depend on the distribution of the hydrophobic, polar, aromatic, small, and other types of residues along the amino acid chain. To assess the peculiarities of distributions of the key amino acids, the frequencies of appearance of pairs of a given amino acid type separated by $i$ number of amino acids ($i=0,1,.....,10$) were estimating and these frequencies were compared with the corresponding frequencies obtained for the randomly generated sequences. Such assessments for the residues of basic (R,K,H) and hydrophobic (V,I,L,F,M) groups are presented on the figures 3a and 3b. Brown middle bars correspond to observed frequencies $fi$ in the set of ribosomal AMPs (relied on the *DBAASP* data (Pirtskhalava et al. 2016). The right dark green and left dark blue bars are corresponding to the values of $Fi + 3\sigma i$ and $Fi - 3\sigma i$ respectively. Where $Fi$ are the average value of the frequencies assessed on the base of random sequences and $\sigma i$ are their standard deviation ($i=0,1,2,...,10$). $Fi$ frequencies were estimated for random sequences generated by the shuffling of sequences of the considered set of peptides. Shuffling is repeated 500 times to assess the average and standard deviation. On the figures 3a and 3b are seen, that an abundance of the pairs of basic amino acids with the 3, 6, 7 residues between them (Fig 3a ) and pairs of hydrophobic amino acids with the 2, 3, 6,7, 10 residues between them (Fig 3b) is not the result of random processes. Such distribution supposes to aim the support the amphipathic alpha-helical conformation of AMPs in the membrane environment.

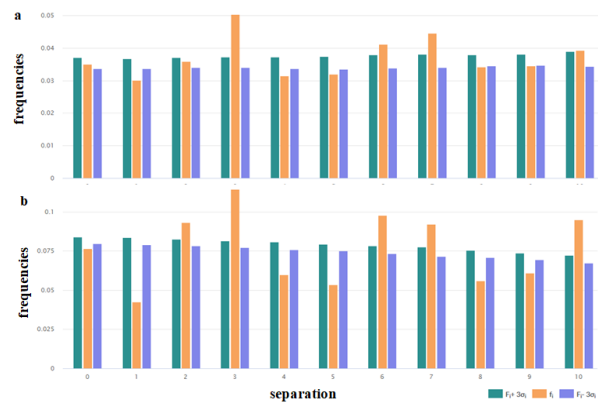

**Fig. 3.** **Frequencies of the appearance of pairs of a) positively charged residues (R, K, H) and b) hydrophobic residues (V, I, L, F, M) with the** *i* **residues (***i*=0,1,2,...,10**) between them**. *fi* - observed frequencies , *Fi*- the average value of the frequencies assessed on the base of random sequences and *σi* are their standard deviations. *Fi* frequencies are estimated for random sequences generated by shuffling of sequences of the considered set of peptides. The assessments have been performed using the tools of the page of "Statistics" (https://dbaasp.org/statistics) of the *DBAASP* (Pirtskhalava et al. 2016)

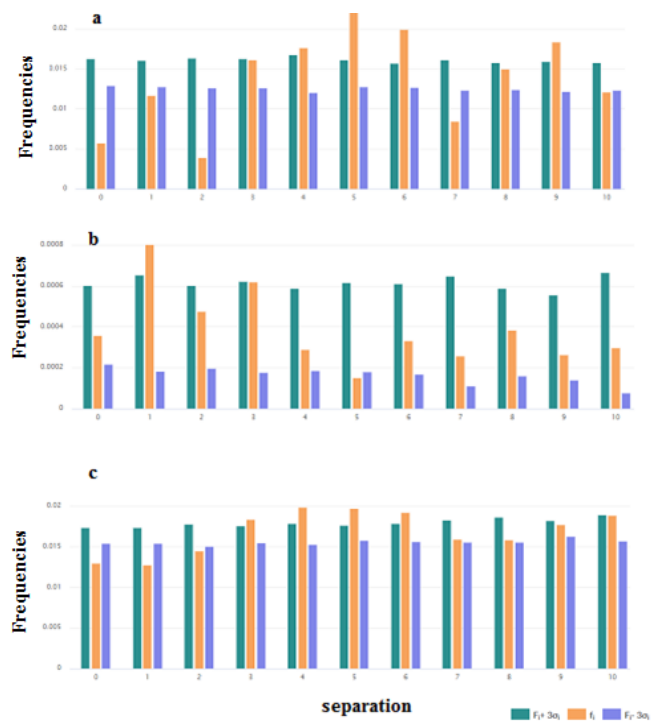

**Fig 4.** **Frequencies of the appearance of pairs of amino acid with the** *i* **residues between them, for: a) Cys (C)** in the disulfide- bonded ribosomal AMPs set**; b) Trp(W)** in the full set of ribosomal AMPs**; c) Gly (G)** in the full set of ribosomal AMPs**;** The assessments have been performed using the tools of the page of "Statistics" (https://dbaasp.org/statistics) of the *DBAASP* (Pirtskhalava et al. 2016)

It was interesting to look for the distribution of the pairs of key amino acids such as Cys, Trp, and Gly. These distributions are shown in Fig 4a,b,c. For the pairs of Cys the distribution was built based on the data of disulfide-bonded AMPs (Fig 4a). From Fig 4a, it's clear that the result about the abundance of the pairs of Cys with the 4, 5, 6 and 9 residues between them is reliable. Such abundance can be connected with the fact that among disulfide-bonded AMPs more than 60% contain only one disulfide bond. The Cystines of such AMPs mainly form loops of about 6 to 9 amino acids long ( including Cys) creating the hairpin or the structure of the shape of a" lasso " (disulfide ring is situated at the one end of the chain).

The preferable distance along the chain in the case of the pairs of Trp is one amino acid (Fig 4b). Aromatic Trp prefers to be in close proximity in the sequences of the ribosomal AMPs.

Gly is the amino acid with the most conformational freedom. Therefore their abundance in the AMPs can be linked with the necessity of conformational flexibility. It's considered that Gly along with Pro are responsible for the creation of the turns and loops in the polypeptide chains to prepare conditions for the interactions between fragments of chain and stabilization of the tertiary structure. Therefore it's easy to find a natural explanation of the abundance of the Gly, Pro, Ser, and Thr in the cyclic AMPs as it was done above. So glycines facilitate formation of tertiary structure that along with a nonvalent interactions can be stabilized by intrachain bonds (including disulfide bonds). Disulfide-bonded AMPs are longer than linear AMPs (Fig 5a,b) and their chains have to bend to form corresponding tertiary structure and Gly in this case plays appropriate role. A length of linear AMPs varies in the interval 10-50 aa. Distribution of the lengths of AMPs indicates on the two major groups of linear peptides: very short with the length (10 – 15) aa, and short with the length (17 - 30) aa. Its shown that in the membrane-bound state majority of linear AMPs have a propensity to the alpha-helical conformation. Are the membrane-bound AMPs' alpha-helices linear or curved? is the question connected with the appearance of the Gly and Pro in certain sites of the chain. It's considered, that Gly along with Pro are responsible for the creation of the kinks in the long alpha-helical fragments of both membrane and soluble proteins (Wilman, Shi and Deane 2014 ). Comparing of the portions of GLy + Pro residues in the very short linear ( length in the interval of 10-15 aa) ribosomal AMPs with the short ( length in the interval of 17-30 aa) ones shows that percentage of GLy + Pro residues in the very short AMPs is lower ( 13%) then in short AMPs (17%).

a)

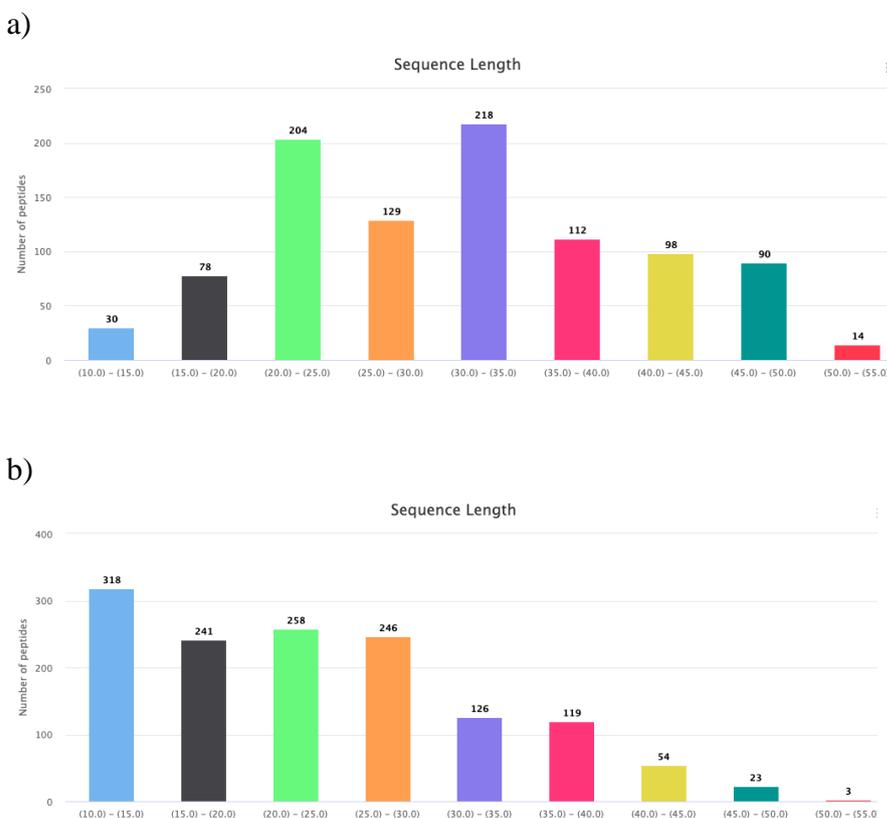

b)

**Fig 5.** AMP's lengths distributions for: a) disulfide-bonded and b) linear peptides (according to the data of *DBAASP* (https://dbaasp.org/statistics))

It's interesting that the distributions of the pairs of Gly-Pro and Pro-Gly are also distinct for both, very short and short AMPs. As it's shown in Fig 6, an abundance of the pairs of the Gly-Pro with the separation of 6, 7 and 10 amino acids points on the concentration of prolines towards C-termini of short AMPs and so, on the convenience to create kinks. In contrast to this the same distributions built for very short AMPs say that prolines are concentrated mainly at the N-termini, to avoid the disruption of the helix.

At the same time, the distribution of pairs of Gly in the set of ribosomal peptides allows supposing that Gly's function is the supporting of the aggregates of AMPs. It's known that GxxxG motives in the alpha-helical fragments of the transmembrane proteins are responsible for the formation of alpha-helical associations in the membrane environment (Russ and Engelman 2000). The results, that show that the abundance of pairs of Gly with the distances of 3,4, 5 and 6 residues along the chain are not random (Fig 4c) can be explained by the demand on the motives necessary to aggregate. Although the kinks can be also considered as a support to raise the capability of the helices to interact (including the formation of helical aggregates ).

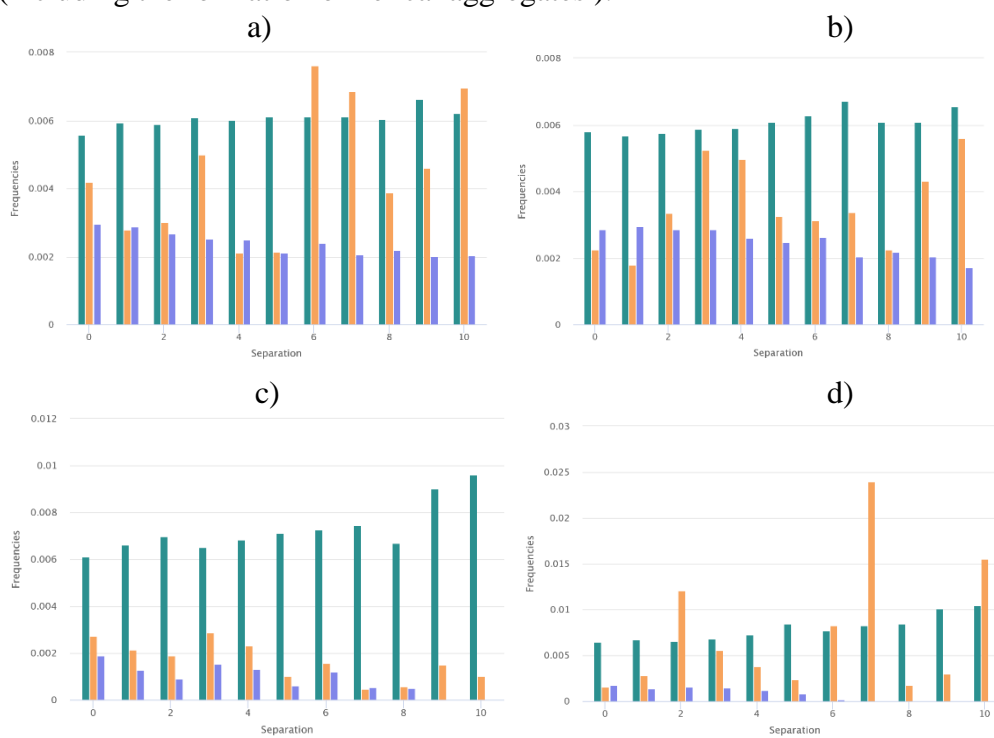

**Fig 6. Frequencies of the appearance of pairs of amino acid with the *i* residues between them, for: a)** Gly-Pro in the set of short (17-30 aa) linear ribosomal AMPs;  **b)** Pro-Gly in the set of short (17-30 aa) linear ribosomal AMPs;  **c)** Gly-Pro in the set of very short (10-15 aa) linear ribosomal AMPs; **d)** Pro-Gly in the set of very short (10-15 aa) linear ribosomal AMPs. The assessments have been performed using the tools of the page of "Statistics" (https://dbaasp.org/statistics) of the *DBAASP* (Pirtskhalava et al. 2016)

So, peculiarities of amino acid composition and their distribution along the chain allow imagining AMPs as the flexible peptides, not aggregative in the water environment, and having the potency to interact with the lipid bilayer (especially with microbial membrane) due to the abundance of the basic and aromatic residues. AMP sequences possess resources to adopt amphipathic alpha-helical conformation due to interaction with the membrane and resources to aggregate in the membrane environment.

### 3.2 Hydrophobicity, Charge, Hydrophobic moment, Isoelectric Point
The distribution of the values of charges of ribosomal AMPs shows that it is closed to Gaussian with average value of 3.41 and standard deviation 2.66 (Fig. 6). AMPs mainly are cationic peptides, although, as has been seen from Fig 7 negatively charged and neutral AMPs, also occur. Worth noting the existence of highly charged ( > +7) peptides, for which mode of action doesn't apparently require the high value of hydrophobicity.

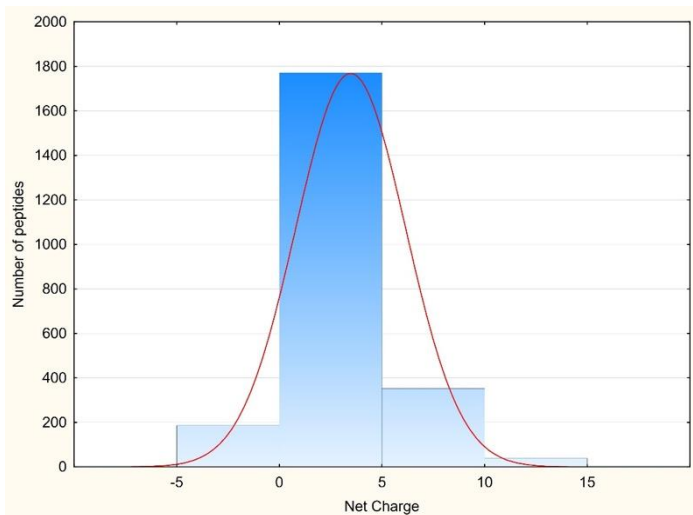

**Fig 7.** Distribution of the charges of ribosomal AMPs using the tools of the page of "Statistics" (https://dbaasp.org/statistics) of the *DBAASP* (Pirtskhalava et al. 2016). The fitting has don by the Gaussian curve (red)

The values of normalized hydrophobicity (assessed relying on Kyte and Doolittle scale (Kyte and Doolittle, 1982)) is also distributed to close proximity to normal distribution and can be approximate by Gaussian function with average -0.25 and standard deviation 0.84 (Fig. 8). Distribution allows declaring that average AMP is a weakly hydrophobic compound, less hydrophobic than trans-membrane peptides (Vishnepolsky and Pirtskhalava 2014; Pirtskhalava, Vishnepolsky and Grigolava 2013). This is the explanation of the fact that the majority of AMPs are functioning in the interface site of the membrane.

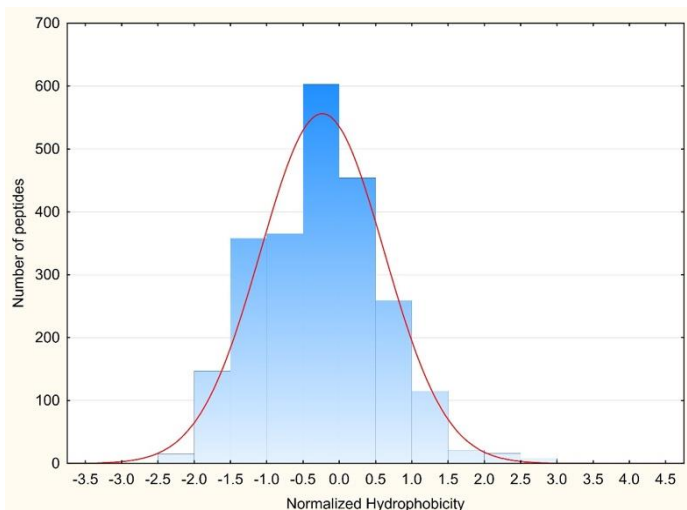

**Fig. 8.** Distribution of the normalized hydrophobicity of ribosomal AMPs using the tools of the page of "Statistics" (https://dbaasp.org/statistics) of the *DBAASP* (Pirtskhalava et al.2016). The fitting has done by the Gaussian curve (red)

It's supposed that the majority of AMPs are amphipathic. To assess the values of amphipathicity, knowledge of 3D structure of peptides is necessary. So we decided to look over the distribution of the amphipathicity in the set of short (8-23 aa), linear peptides the structure of which in membrane environment is reasonable to consider as alpha-helical (Fig.9). Distribution of *normalized hydrophobic moment ($\mu$)* can be approximated with sum of two Gaussian functions. The distribution allows to suppose existence of two type of peptides: more amphipathic with average $\mu$ = 1.5 ($\pm$ 0.35) and less amphipathic with average $\mu$ = 0.4 ($\pm$ 0.3).

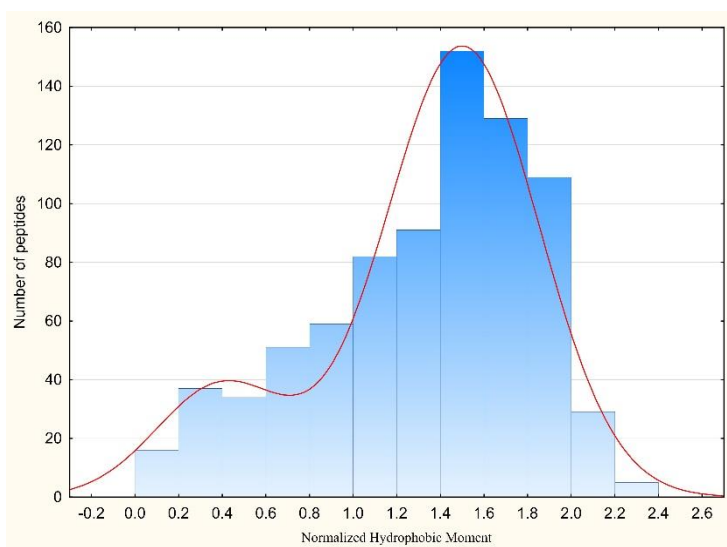

**Fig. 9.** Distribution of the normalized hydrophobic moment ($\mu$) of short (8-23 aa), linear, ribosomal AMPs using the tools of the page of "Statistics" (https://dbaasp.org/statistics) of the *DBAASP* (Pirtskhalava et al. 2016). The fitting has done by the sum of two Gaussian curve (red)

It can be proposed that these two types of AMP behave differently in the membrane, that is their mods of action are different.

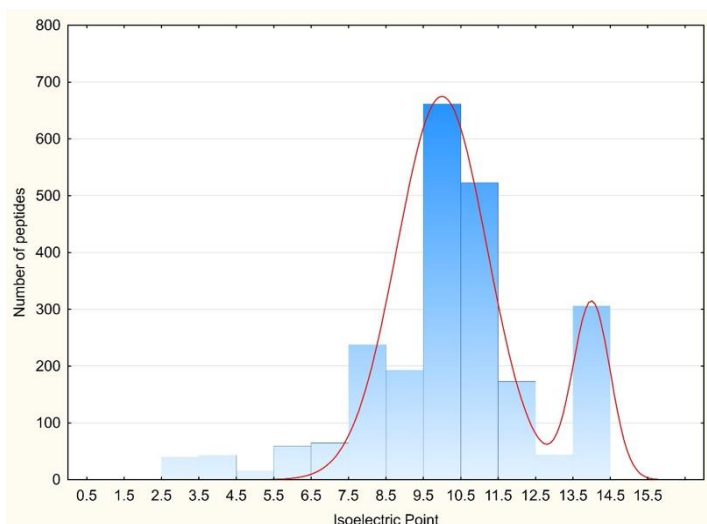

**Fig. 10.** Distribution of the Isoelectric Point of ribosomal AMPs using the tools of the page of "Statistics" (https://dbaasp.org/statistics) of the *DBAASP* (Pirtskhalava et al. 2016). The fitting has done by the sum of two Gaussian curves (red)

The distribution of the values of Isoelectric Points of ribosomal AMPs(Fig. 10) also points on the existence of the two types of peptides with average values and standard deviations of 10$\pm$1.2 and 14$\pm$0.5, respectively.

## 4. Secondary structure and self-aggregation

According to the 3D structure, AMPs are classified as alpha-helical, beta-structural, alpha +beta , unordered, etc. But this classification conditional, because structure of many AMPs is determined by the environment and changed depending on the variation of environment and/or other factors (a local concentration of peptide for instance).

In membranes, the rules governing the formation of secondary structure and folds of the polypeptide chain are very different from the aqueous environment. In the membrane, hydrogen bonds become probably more important for driving secondary structure and their aggregates formation. The hydrogen-bond effect helps explain the easy formation of secondary structure in membranes (Kaiser and Kézdy 1983). Well known, that each amino acid has its own propensity to a particular secondary structure. The propensity of individual amino acids to a particular secondary structure may be altered in response to the change in the environment from aqueous to the membrane.

In the aqueous environment, the bulky aromatic residues (Tyr, Phe, and Trp) and β-branched amino acids (Val, Ile) are favored to be found in β strands in the middle of β sheets. Glycine is an intrinsically destabilizing residue in β sheets. In natural proteins, however, this destabilization can be compensated by specific cross-strand pairing with aromatic residues (Merkel and Regan 1998 ). At the same time, a β-branched residue, such as Ile and Val are described as α-helix destabilizing. Gly and Pro also destabilize the α-helices in globular proteins.

In the membrane environment Ile and Val rank as the best "helix-promoters" and it was found that they be important for membrane protein assembly and folding (Li and Deber 1994 ). Gly and Pro also display a considerable tendency to form α-helices in membrane environments (Li and Deber 1994 ). The abundance of bulky amino acids Phe, Trp, Ile, Leu in linear AMP tend to promote α-helix formation in the membrane by interaction with aliphatic chains of the bilayer's core, while at the more polar interface area they can stabilize beta structural aggregates.

In the aqueous environment bulky residues and positive charges can block alpha helical self-aggregation. The formation of beta structure can be also inhibited by abundance of positively charged Lys. Therefore majority of linear cationic AMPs are disordered in water environment and adopt regular secondary structure after interaction with membrane. Here it has to be noted, that AMPs rich of basic amino acids and/or prolines have preference to ppII conformation (Shi, Woody and Kallenbach 2002.) Cyclic and majority of disulfide-bonded peptides form a well-defined structure in solution. But their structure in the membrane is membrane-dependent. For instance, the PG-1 forms oligomeric transmembrane β-barrels in bacteriamimetic anionic lipid membranes, whereas in the cholesterol-rich membranes mimicking eukaryotic cells the peptide forms β-sheet aggregates on the surface of the bilayer (Tang and Hong 2009 ).

**4.1 Secondary Structure**

Amino acid composition allows making a supposition on secondary structure and propensity to an aggregation of AMP. Secondary structure and a propensity to aggregation along with the hydrophobicity and charge are essential determinants of antimicrobial potency. Peptide secondary structure and its propensity to aggregation depend on environment. So, AMPs, mainly linear, can adopt different conformations depending on environment. The conformation of peptide GL13K supposed to be disordered in water, α-helical in the zwitterionic lipid bilayer and beta-structural in an anionic lipid environment, especially when peptide to lipid ratio is high (Harmouche. *et. Al.* 2017). At high local concentration predisposition of peptides to self-aggregation into beta structure rises. So, the GL13K−membrane interactions are governed by an equilibrium between the random coil, α-helical, and β-turn conformations. At the high concentration, GL13K can initiate the β-sheet aggregates. Some other peptides ( Catestatin (Sugawara *et al.* 2010), cateslytin (Jean-Francois *et al.* 2008), fusion peptide of HIV (Lai *et al.* 2012) similarly has exhibited a varying equilibrium between α-helical and β-turn secondary structures depending on lipid composition.

The classic example of peptides that undergo a conformational transition from the native, mainly α-helical structure into an isoform with high beta-sheet content, are amyloid-forming peptides. They share key structural and functional features with AMPs. Although some amyloid-forming peptides possess antimicrobial potency (Kumar *et al.* 2016), they are well studied mainly because considered as immediate precursors for the formation of amyloid fibers and so, as the most toxic components in many neurodegenerative diseases. To unravel the molecular interactions that occur during the transformation from alpha-helix to beta-sheet, the model peptide has been designed based on the well studied α-helical coiled-coil folding motif (Pagel *et al.* 2006). Study of the model peptide shown that the resulting secondary structure is strongly depended on environmental parameters. Worth noting, that peculiarities of transition between secondary structures predetermines a velocity of formation and type of aggregates caused by self-assembling (Harmouche. *et. al.* 2017; Pagel *et al.* 2006 ).

One more type of secondary structure that might be considered for AMPs is ppII. ppII conformation is convenient for the peptides rich with basic amino acids and/or prolines. Buforine, cell-penetrating peptide, due to propensity to the ppII conformation finds its cytoplasmic target, DNA (Lan *et al.* 2010).

**4.2 Self-aggregation**

Association of AMP with the lipid bilayer and creation of lipid-peptide complexes is a necessary step at the action on to the microbial cell. Along with the ability to aggregate with lipids, a propensity to self-aggregation is crucial for the capability of AMP to affect microbial membrane (Sarig *et al.* 2008).

Although not all-AMPs can self-assemble, it has been considered as an important property of peptides, because can be influential to the potency and mode of action. Peptides self-assembly is driven by electrostatic forces, hydrogen bonding, hydrophobic interactions and π-π stacking interactions (Bowerman *et al.* 2011). Consequently, the propensity to aggregation depends on environmental conditions.

Because AMP reaches target membranes through the aqueous phase, their properties in aqueous solutions are important for their effects on membranes. Rina Feder and coworkers tried to understand how AMP organization in aqueous solution might affect the antimicrobial activity and conclude that potency correlated well with aggregation properties. They have shown that aggregation can have dramatic consequences on the antibacterial activity of the Dermaseptin-derived peptides. More potent against bacteria peptides were clearly less aggregated (Feder, Dagan and Mor 2000 ) in a aqueous environment. Developing predictive models by the artificial neural network it has been found that peptide aggregation in solution indeed contributes to low antimicrobial activity (Torrent et al. 2011a). Interestingly, the addition of cationic residues to peptides has been shown to inhibit aggregation in solution while improving the antimicrobial potency at the same time (Torrent et al. 2011b).

The propensity to aggregation is determined by sequence and consequently structural features of AMP. Linear AMP, because of its short length, a combination of non-high mean hydrophobicity with relatively high net charge shows the disordered structure in aqueous solution and prevention of the aggregation (high net charge and the resulting electrostatic repulsion between peptides limits aggregation). Cycled main chain or intra-chain covalent bonds represent a prerequisite for certain structural stability of AMP in aqueous solution and for the possibility of aggregation.

Self-assembling of linear peptides on the cell membrane can contribute to antimicrobial potency. It's suggested that either in the pore formation or detergent-like mechanism, the process of the self-assembling of AMP is involved (Bechinger 2015; Sato and Feix 2006). Recently an attempt has been made to demonstrate a functional relationship between the AMP self-assembly and their antimicrobial activity (Ye *et al.* 2019). As above mentioned, the self-aggregation of cationic AMP is mainly governed by electrostatic interactions. Differences in the behavior of AMP depending on environment (aqueous or membrane) are explaining by the weakening of repulsion between positively charged groups in membrane. An attempt (Ye *et al.* 2019 ) has been made to look over the behavior of designer peptides GL13K and their variants, including D-enantiomer at various pH of the solution, to model the process of weakening of repulsion. The study of structural links between secondary structure, supramolecular self-assembly dynamics, and antimicrobial activity has been performed. It's not surprising, that variation in pH and consequential evolution of secondary structures were related to a self-assembly process. It's interesting that the time of the initiation of self-assembly determines antimicrobial potency. For instance, two GL13K enantiomers formed analogous self-assembled structures, but D-GL13K initiated self-assembly faster and had notably higher antimicrobial potency than L-GL13K.

It's considered that the intrinsic antibacterial capabilities of peptide-based supramolecular assemblies have been largely overlooked and need more attention. The antibacterial activity of self-assembled diphenylalanine has been studied in order to gain insights into the significance of the interplay between selfassembly and antimicrobial activity (Schnaider *et al.* 2017). Worth noting, that diphenylalanine is the central recognition module of the β-amyloid peptide. The study demonstrated, that the interaction of diphenylalanine with bacteria causes damage to bacterial morphology, especially membrane and thereby inhibiting bacterial growth. The non-assembled diphenylalanine samples, which consisted of the peptide at sub-critical concentrations, inhibited bacterial growth by 15–20%, only. Results underline the significance of self-assembly to the antimicrobial activity and allow to announce, that diphenylalanine motif can be used as a minimal self-assembling antimicrobial building block and due to its low cost and high purity can serve as an important platform for the development of alternative antimicrobial agents and materials (Schnaider *et al.* 2017 ). Hydrophobic and non-cationic nature of diphenylalanine makes it an attractive compound to combat resistance, because, as known several bacterial strains are developing countermeasures mainly by modification of cell envelope, reducing an electrostatic attraction to the membrane.

Worth-noting, that along with the hydrogen bonding, hydrophobic and electrostatic Coulombic interactions it is suggested that self-assembly can be governed by aromatic π-π interactions. Although a role for aromatic π-π interactions in peptide self-assembly is the object of debates. Relying on the results of a study of model peptides Ac-(XKXK)(2)-NH(2), where X=Val, Ile, Phe, pentafluoro-Phe and cyclohexylalanine, has been concluded that aromatic amino acids do not always more readily induce self-assembly relative to nonaromatic amino acids of similar hydrophobicity. At the same time, aromatic amino acids can create unique morphology of fibril (Bowerman *et al.* 2011 ).

Consequently, for some AMP an interplay between self-assembly and secondary structure in the membrane environment is crucial for the action on the membrane and to emerge antimicrobial potency. At the same time, the mode of actions of some other AMPs don't require self-aggregation, other properties predetermine their antimicrobial potency.

## 5. Frequently occurred amino acids

Among AMPs the peptides rich of particular amino acids, such as Arg, Trp, Pro, Gly, Cys, His, etc occur. Modes of action of such peptides are determined by the physicochemical features of these residues. We will try to overview the knowledge of the behavior of such peptides in the membrane environment.

### 5.1 Basic amino acids and AMPs

It's known that AMP is majorly cationic peptides. But even among cationic AMPs a peptides with a high content of basic amino acids (percentage of basic amino acids 20-30% or more) occur. Such peptides are not self-aggregated in the polar (aqueous) environment, and it has been supposed that some of them (percentage of basic amino acids >50% ) due to low amphipathicity cannot self-aggregated even in the membrane. Because of such extremes, it's reasonable to suppose for the peptides with a high content of basic amino acids (HCAMP) a peculiar, different from other AMPs, mode of action.

**Lysine- and Arginine in the AMPs.** The mode of action of HCAMP is reasonable to consider in the light of data gained for cell penetrating peptides (CPPs) because HCAMP and CPP shared physicochemical features and CPPs are membrane active peptides also. The peptides: Penetratin, Tat, Arg9, R6/W3 considered as CPPs and at the same time possess an antimicrobial potency (Di Pisa, Chassaing and Swiecicki 2015 ). Moreover, the energy-independent mechanism, named "direct translocation" has been fully characterized for last peptides. Hallmark of CPPs is an abundance of basic (Arg and Lys) residues and/or Trp. Study of internalization process of these peptides shown, that Penetratin and R6/W3, which are amphipathic, preferentially interact with anionic regions of membrane, while not-amphipathic Arg9 and Tat interact only with anionic lipids. Interaction with anionic headgroups of lipids seem to be sufficient for Penetratin to trigger a helical conformation (Magzoub, Eriksson and Gräslund 2002), which in water is unordered. However, for a higher content of anionic lipids, the α-helicity decreases in favor of β-sheet structures( Eiríksdóttir *et al.* 2010 ). In contrast to Penetratin , Tat and Arg9 remain unfolded even in the membrane environment. So, it's supposed that direct translocation of R6/W3, and Penetratin occurs when they are in a helical conformation, whereas Arg9 and Tat translocate as random coils. The studies of CPP translocation exhibit the crucial role of basic amino acids, especially Arg for the last process. It's shown, that translocation depends on the membrane potential also.

Although both, Arg and Lys are basic amino acids, with a high amphipathic index (Mitaku, Hirokawa and Tsuji 2002 ), they are differently interacting with cell membranes. Guanidinium group of Arg is a major cause of this distinction. It's shown that the guanidinium group binds to the phosphate group of lipids more strongly than the amino group of Lys does this (Robison *et al.* 2016). Moreover at the binding state guanidinium group polarity reduced and it possesses a capability to internalize membrane and to come to close contact with another guanidinium group to form a stacking contact (Vazdar, Uhlig and Jungwirth 2012; Schwieger and Blume 2009 ). These features of the guanidinium group provide penetrating capability of Arg9 peptide, whereas Lys9 does not display a propensity to translocate across membranes (Mitchell *et al.* 2000). Both Arg 9 and Lys 9 show anti-cooperative binding with membrane constituted of a mix of POPC and POPG lipids at low (up to 10%) POPG concentration, although Arg9's anti-cooperativity is weaker. When concentration of POPG rise up to 30%, Arg9 binding becomes non-cooperative, while Lys9's binding remains anti-cooperative (Robison et al. 2016 ). It's shown, that Arg9 can form stable complexes with artificial membrane constituted of zwitterionic lipid POPC, while Lys9 can not (Robison et al. 2016 ). These peculiarities of side chains of Arg and Lys, allow supposing that an abundance of Lys in ribosomal AMPs connected with the necessity to rise a selectivity of AMPs to the prokaryotic cell membranes. Abundance of Lys in ribosomal AMPs indicates different from penetrating peptides modes of action for the majority of ribosomal AMPs. Although Arg-rich peptides appear among ribosomal AMPs also.

MD simulations of TAT peptide in the model membrane constituted of zwitterionic lipids shown, that a basic amino acids of cationic peptides try to move close to the phosphate group of lipids of both proximal and distal layers of membrane (Herce and Garcia 2007). At the high peptide to lipid ratio, attached to surface peptides produce a thinning of the membrane and so reduce the hydrophobic free-energy barrier to allow some side chains of basic amino acids to reach phosphate group of the distal layer. The length of the lysine and arginine side chains facilitates to do this. As a charged group enter the hydrophobic lipid bilayer, water followed them. The water and phosphate penetration create transient pores and so allow some other peptides diffuse across them (Herce and Garcia 2007). It has to be noticed, that the Arg insert into the membrane more easily than Lys (Wender et al. 2000 ). As above mentioned, the special binding of the Arg to phosphate groups is explained by the specific structure of the guanidinium group.

Thus, electrostatic interaction can be considered as a driven force at CPP translocation across the membrane. Consequently membrane potential has to affect the process of translocation. To explore an impact of membrane potential on CPP translocation another MD simulation (Gao et al. 2019) has been performed for four types of peptides (WALP carrying zero, INLK carrying three, TAT carrying eight and R9 carrying nine positive charges). Three lipid components, including DPPC, POPC, and cholesterol were used to build a one-bilayer or two-bilayer model of the membrane. The local membrane potential has been produced by the ion concentration imbalance across the membrane or by external electric fields. Atomistic molecular dynamics simulations demonstrated that local membrane potential plays an essential role in translocating process. Due to enhanced local membrane potential, CPPs readily enter cells through the opened membrane pore in a chain-like conformation. Penetration time was consistent with experimental data (Gao et al. 2019 ).

CPP is mainly cationic peptides and we see that translocation across the hydrophobic part of the bilayer is not passive diffusion. CPP- membrane interactions cause some perturbations in the arrangements of lipids in the membrane. For the microbial membrane, where a portion of anionic lipids is relatively high and so attraction of peptides to the membrane is strengthened, more marked perturbations can be supposed. The prokaryotic membrane will promote an accomplishment of the high local concentration on their surface because it can provide the more sharp partition of peptides between the aqua and membrane. So, for HCAMP, the model that intends a perturbation of lipid bilayer with the creation of transient pore is even more acceptable in the case of the microbial membrane. Here worth noting, that membrane potential, which is correlated with translocation capability of HCAMP is markedly higher in the case of the prokaryotic cell than eukaryotic.

Peptide to membrane binding can be the cause of structural changes of the membrane and the carpet model is a widely used model describing the lipid packing defects induced by CPPs (Di Pisa, Chassaing and Swiecicki 2015). But some other modifications induced by the CPPs are considering also. For instance, in the artificial membranes constitute of zwitterionic and anionic lipids (DMPC/DMPG) have been shown that Penetratin induces membrane invaginations resulting in the formation of tubular structures. Moreover, has been demonstrated that membrane fluidity was crucial in the occurrence of membrane deformations after Penetratin binding. In a membrane of fluid disordered phase (Ld), Penetratin is able to induce invaginations. At the same time, the peptide doesn't have an effect on a raft-like membrane (Lamaziere et al. 2008). So, exploration of the interaction of CPP with different membrane domains and revealing of the modifications of lipids' organization are important for understanding the mechanisms of translocation. The influence of the CPP on the different microdomains of the plasma membranes with different phospholipid compositions has been investigated (Almeida et al. 2016). The results indicate that Penetratin is able to induce rearrangements of membrane lipids that favour phase separation and membrane heterogeneity (Walrant et al. 2012). R6/W3 and Arg9 also have capability to increase membrane fluidity that facilitate peptide translocation (Walrant et al. 2012; Walrant et al. 2011).

**Histidines in the AMPs.** Because the isolated Histidine, has a pK of approximately 6.5 it's largely unprotonated and uncharged at physiological pH while are protonated and cationic at acidic pH. So, histidine behaves as a basic amino acid at the acidic pH only. At the same time, Histidine is classified as aromatic due to the presence of a sextet of p-electrons. Consequently, Histidine can be involved in the different types of interactions: the coordinate interactions between histidine and metallic cations are the strongest, followed by the cation-π, hydrogen-π, and π-π stacking interactions (Liao et al. 2013). When the histidine is in neutral form, the cation-π interactions are attractive; when it is protonated, the interactions become repulsive. The complicated nature of Histidine predetermines the pH-dependent mode of interaction of the His-rich peptides with the membranes (Kacprzyk et al. 2007). At physiological pH, His -rich peptides behave differently from Arg- and Lys-rich peptides, while at the lower pH they resemble Arg-rich peptides. For instance, histidine can form stable noncovalent complexes with acidic residues, including a phosphate group. These interactions have a similar chemical basis as in the case of Arg although they are weaker (Muller, Jackson and Woods 2019). Exploration of the empirical distribution of the protein-ligand cation–π interactions found in X-ray crystal structures shows that positively charged histidine residues are rarely involved into cation–π interactions, although the stacked π+–π interaction is estimated to be of similar magnitude to that of arginines (Kumar et al. 2018). By analogy to Arg, a protonated histidine has a propensity for forming like-charged contact pairs with another protonated histidine or with arginine (Heyda, Mason and Jungwirth 2010)

A pK of the histidine side chain in the peptide depends on the local environment. It's assessed, that the pKs of histidine side chains of the SDS bound Hb-33–61peptide were on the order of 7.7 to 7.8 (Sforça et. al. 2005 ). Hb-33–61 is a proteolytic product of the bovine hemoglobin alpha-chain and has potency against a Gram-positive bacteria and fungi. It's supposed, that small changes in the local pH may have large effects on the activity of histidine-containing AMP. For instance, the cationicity of the clavinines , other histidine-rich peptides produced by the solitary tunicate Styela clava, derives primarily from histidines, and they are active at pH 5.5 but relatively inactive at pH 7.4 (Lee, Cho and Lehrer 1997). Moronecidin, the 22 residues long peptide by analogy to the clavinine contains 4 histidines but in addition other basic residues also. The greater positive net charge of moronecidins accounts for the antimicrobial activity at neutral pH (Lauth et al. 2002). Consequently, the pH-responsive activity changes depend not only on the portion of histidines in the peptide but on the overall composition and position of the histidines in the chain. Histatin peptides belong to a family of salivary histidine-rich AMPs. Histatin-1 and Histatin-3 are derived from the humans' genes HTN1 and HTN3 (VanderSpek et al. 1989). Histatin-5 is a proteolitic fragment of the histatin-3. Despite the fact that histatins exhibit a high degree of sequence homology, histatins 1 and 3 's activities are pH-dependent, while the activity of histatin 5 is pH-independent over the range of pHs 4 to 8. Differences in the behavior of histatins explained by the acidic residues present in the carboxylterminal domains of histatins 1 and 3, which are absent in histatin 5 (Xu *et al.* 1991).

The mode of action of histatins is a subject of intense debate. It's supposed that all targets of histatins are intracellular. Translocation into cell may involve different uptake pathways (energy -dependent and energy-independent). There is a suggestion, that once inside the fungal cells, Hst 5 affect mitochondrial functions (Puri and Edgerton 2014 ). As above mentioned,  His-rich peptides bind copper (Cu) and other metal ions in vitro. Bis-His motifs are commonly found in biological systems that can form a coordination motif to create Cu(I)−bis-His complexes with O2 reactivity. This allow to speculate that the Cu(I)−histatin complex could potentially mediate Cu-induced oxidative stress, which in turn  may affect mitochondrial functions and could be the cause of the fungal killing. It's proved that Cu modulates histatins' antifungal activity (Conklin *et al.* 2017). By analogy, the $Zn^{2+}$ ions are modulated antimicrobial activity of Clavanin A (Juliano et al. 2017 ). $Zn^{2+}$-dependent antibacterial activities were shown for other histidine –rich peptides (Kacprzyk *et. al.* 2007 ), including histatin 5.  Worth noting that,  $Zn^{2+}$ is considered to be functionally similar to the acidic pH, which impose a positive charge on histidine-rich peptides (Rydengård, Andersson Nordahl and Schmidtchen 2006). It's shown by confocal microscopy, that Clavanin A -Zn2+ added to Escherichia coli translocate across the cell membrane to find a cytoplasmic target (Juliano et al. 2017 ).

To their features, Histidine is considered as a convenient residue to tune pH- and cationic ion-sensing, cell-penetrating sequences. The family of Histidine-rich designer peptides, LAH4 is a well-studied example of such peptides.  The development of LAH4 peptides was aiming to solve the transfection problem, but they possess antimicrobial activities also. Both processes, transfection and antimicrobial action require membrane-active sequences. For transfection, the peptide has to be involved in the trafficking to the endosome and when inside the cell, to provide an escape from the endosome.  So the interaction of LAH4 peptides with the membrane was majorly studied on the model vesicles mimicking plasmatic or endosomal membranes of the eukaryotic cells (Wolf et al. 2017).  Both antimicrobial activities and endosomal membrane disruptive capabilities of LAH4 peptides are pH-dependent. Studies of action of LAH4 peptides on the vesicles mimicking the prokaryotic membrane, that is membrane constituted of anionic POPG lipids, shows that calcein release activity depends on pH and acyl chain s' saturation (Vogt and Bechinger 1999; Perrone *et al.* 2014) . By adding additional residues being positively charged at the neutral pH, the HALO family of peptides active even at neutral pH has been created. (Mason et al. 2009 ). It should be noted, that although the activity and pH selectivity of histidine-containing peptides are mainly determined by the number of histidines, peptides with the same number of histidine at different positions give neither the same activity nor the same pH sensitivity profiles.  Therefore, it can be tuned the pH-sensitivity of histidine-containing peptides by manipulating histidine numbers and positions (Tu  et  al. 2009) to design agents that would function selectively in acidic compartments.

### 5.2 Aromatic amino acids and AMPs.  Cation- π  interactions
It's known that aromatic systems interact strongly with cations. The side chains of Phe, Tyr, and Trp in the proteins are considered as a cation-binding site, a so-called "hydrophobic anions" (Dougherty  1996 ). Because of the indole ring, tryptophane provides a much more intense region of negative electrostatic potential than does benzene or phenol, and so in proteins, it appears more intensively at the cation-π interaction sites.

The complicated nature of Trp side chains predetermines its behavior in the membrane environment also. A membrane -Trp interaction is complex and governed by hydrophobic effect, dipolar, quadrupolar, H bonding, and cation-π interactions. Hydrophobic effect drives Trp out of the water, while complex electrostatic interactions push it to the headgroups of the lipids (to the hydrated interface of the membrane) (Yau *et al.* 1998). Distribution of individual amino acid in known structures of helical membrane proteins has been studied, and propensities of their occurrence at different portions of the bilayer as a function of depth in the bilayer were calculated (Senes *et al.* 2007). The function that describes the behavior of Trp (a propensity to particular membrane site) corresponds to Gaussian distribution with the global maximum at the distance of 11.9 Å from the middle plane of membrane. So, an estimation of how deeply a side chain of Trp prefers to penetrate into a membrane shows that preferable depth of penetration corresponds to the interfacial region of the membrane. Sigmoidal curves describe the behaviour of other bulky hydrophobic amino acids, such as Leu, Ile, and Val, showing that they penetrate into membrane more deeply.

A high content (relative to average protein) of aromatic hydrophobic residues in the linear ribosomal AMPs allows suggesting an essential role at the functioning. Moreover, there is a group of AMP where Trp and Arg have a prevalence relative to other amino acids. The effort has been spent to understand the mode of behavior of Trp rich peptides in the membrane and to look for the causes of the differences from the peptides rich with other bulky hydrophobic amino acids. For instance, RW9 and RL9 peptides that have the same length and charge and similar hydrophobicity (according to the Eisenberg scale (Eisenberg 1984)) behave differently in the lipid bilayer (Walrant *et al.*2013). RL9 is more deeply inserted into membrane than RW9, though RW9 can translocate across the membrane in contrast to RL9.

AMPs majorly are cationic peptides and we above saw how can Arg drive the interaction of the peptide with the membrane. The abundance of Trp with its complicated nature put additional capabilities at the interaction of AMP with the membrane. Trp is considered hydrophobic due to its uncharged side chain, while do not reside preferably in the hydrocarbon region of lipid bilayers (Senes *et al.* 2007). Another important property of Trp is the extensive π–electron system of the aromatic indole sidechain that gives rise to a significant quadrupole moment (Dougherty 1996 ). Consequently, basic residues and Trp are capable of participating in cation–π interactions, thereby facilitating enhanced peptide–membrane interactions. The cation–π interaction can take place in either a parallel (stacked) or a perpendicular orientation. In the stacked conformation, the Arg side chain is able to form the same amount of hydrogen bonds as when it is not involved in cation–π interactions (Aliste, MacCallum and Tieleman 2003). This is in contrast to lysines, which cannot form hydrogen bonds while participated in cation–π interactions. The stacked arrangement between Arg and Trp residues is preferred in contrast to Lys and Trp. The cation–π interaction possibly restraining the peptide structure in a suitable conformation to interact with the bacterial membrane. Arg and Trp rich peptides can lead to structures that go far beyond regular α-helices and β-sheets. The Arg is effectively shielded from the highly hydrophobic nature of the bilayer by associating with a Trp residue when peptide penetrates into bilayer (Dougherty 1996; Jing, Demcoe and Vogel 2003).

Indolicidin is a well-studied, short, 13 amino acid Trp-rich antimicrobial peptide. It belongs to the cathelicidin family of peptides. The proportion of Trp residues in Indolicidin is the highest (about 40%). The peptide is unordered in water and adopts wedge-type shape in the micelles (Rozek, Friedrich and Hancock 2000 ). Trp residues are segregated from positively charged ones and situated in a trough, between positively charged regions. It's shown that Indolicidin can cross the membranes at concentrations above the MIC but below the minimal lytic concentration (Hsu et al. 2005).

Antimicrobial peptides of the family of Lactoferricins that known as Trp rich peptides, show cell-permeable capabilities. Interacting with the membrane via electrostatic and/or hydrophobic interactions they may form pores or inverted micelles to shuttle inside the cell (Joliot and Prochiantz 2004). Their mechanism to enter cells is similar to CPPs. Once in the cell, they can interact with DNA or RNA affecting their synthesis.

Puroindolines are small, cationic proteins. They contain Trp rich regions (Blochet et al. 1993). Mode of action of the 13-residue Trp –rich fragment of puroindoline, the puroA has comprehensively studied by a variety of biophysical and biochemical methods (Jing, Demcoe and Vogel 2003). In the bound to membrane state all the positively charged residues are oriented close to the face of Trp indole rings. Due to the high content of Trp residues, puroA is located at the interface site of membrane. The binding has an impact on the phase behavior of the vesicles. The amphipathic structure that appeared upon binding allows the peptide to insert more deeply into bacterial membranes and perturb the membrane bilayer structure (Jing, Demcoe and Vogel 2003). The penetration of puroA into vesicles resembling bacterial membranes was more extensive than into vesicles mimicking the eukaryotic membrane.

**5.3 Prolines and AMPs. ppII conformation.**
Proline is $\alpha$-helix breaking residue. Helix kinks are a common feature of proteins. They raise the conformational flexibility of the helical fragments (Wilman, Shi and Deane 2014). It's shown that proline residues in natural antimicrobial peptides define a hinged region that is crucial for antibacterial potency and selectivity (Vermeer et al 2012 )

When the portion of the Pro in the AMPs is very high, as in the Pro-rich antimicrobial peptides (PrAMPs), the mode of action does not involve the lysis of bacterial membranes but only penetration into susceptible cells, where PrAMPs act on intracellular targets (Scocchi, Tossi and Gennaro 2011).  PrAMPs are a group of cationic host defense peptides of vertebrates and invertebrates characterized by a high content of Pro-s, often associated with Arg-s in repeated motifs. PrAMPs show a similar mechanism and selectively kill Gram-negative bacteria, with low toxicity to animals.

Drosocin, pyrrhocoricin, and apidaecin, representing the family of short (18-20 amino acids) Pro-rich antibacterial peptides, originally isolated from insects and act on a target bacterial protein chaperone DnaK (Otvos et al. 2000). The Pro-rich peptides have multiple functions and functional domains and perhaps carry separate modules for cell entry and bacterial killing. The Pro-Arg-Pro or similar motifs assist the entry into bacterial cells without any potential to destabilize the host cells, and therefore without toxicity to eukaryotes. The antibacterial activity of the native products is provided by the independently functioning active site, capable of binding to the bacterial DnaK and preventing chaperone-assisted protein folding (Kragol et al. 2002). Pro-rich cell penetration modules may be general for antibacterial peptides in nature. For instance, the cathelicidin hydrophobic tail sequences reveal strong similarities to C terminal tails of pyrrhocoricin, drosocin or apidaecin (Kragol et al. 2002).

Uversky et al. have shown that the combination of low mean hydrophobicity and relatively high net charge is an important prerequisite for the absence of stable structure in proteins under physiologic conditions, thus leading to "natively unfolded" proteins (Uversky, Gillespie, and Fink 2000). Then the view has been offered, that unfolded peptides and proteins have a strong tendency to poly-proline II (PPII) conformation locally while conforming statistically to the overall dimensions of a statistical coil (Shi, Woody and Kallenbach 2002). The PPII helix is often observed in the context of Pro-rich sequences, but sequences that are not enriched in Pro can adopt this structure also. It has been shown, Arg9 and TAT peptides, which are arginine-rich cell-penetrating peptides adopt PPII type conformations in the membrane. By analogy to Pro-rich peptides, the arginine-rich peptides are also able to interact with biological membranes and form the transient pores only, without lysis of cells (Herce et al. 2009)

It can be suggested, that linear, cationic AMPs which are disordered in aqueous solution and thought to be in a statistical coil state may, in fact, be flickering in and out of a metastable PPII helical conformation. In the membrane environment, many of them form high ordered structures (alpha-helical, beta structured or even aggregated) and so cause membrane disturbance and permeabilization. But some of them, for example, Pro-rich peptides do not behave so, as they are not capable to form high order structures in the membrane and remain in ppII conformation in the latter environment. Proline induces the conformational flexibility of the polypeptide chain and so, prevents a self-association of peptides. Flexibility is crucial for antibacterial potency and selectivity of many AMPs (Vermeer et al. 2012 ). The proline residue in Buforin II, for instance, operates as a major translocation factor (Park, Kim and Kim 1998). To analogy to Buforin II, pro-rich peptides show membrane penetrating capability and find targets for antimicrobial functioning inside a cell, with which form stable complexes. In this respect, PrAMPs resemble Arg rich penetrating peptides.

It should be noted, that charged residues and Pro is the worst aggregators in both alpha and beta self-aggregation (Pawar et al. 2005). The pro-rich peptide is capable to aggregate only with conjugated to Gly form as it takes place in the case of collagens or other Pro/Gly rich polypeptides (Creasey, Voelcker and Schultz 2012). Collagen's chains adopt conformation somewhat similar to ppII. Worth noting that ppII conformation and not $\alpha$–helical is convenient to the aggregation with DNA (Lan et al. 2010)

**5.4 Glycines and AMPs**
Glycine is the amino acid with the highest conformational freedom. So their appearance in the sequence can be connected with the necessity to raise the flexibility of the peptide. Another hallmark of the Gly is a small side chain which facilitates the formation of the $C^{\alpha}$- H hydrogen bond. These two features promote the unexpected high content of Gly in the helical fragments of transmembrane proteins. Moreover, the majority of Gly from transmembrane helix situated in the hydrophobic area of the membrane and supports the stability of the complexes of the transmembrane helical fragments (Dong et al. 2012). The support is expressed in the promotion of the inter-helical hydrogen bonds and/or by creation kink in the helix. To create pro-kink area Gly frequently acts in the pair with Pro. Such a pair weakens single helix stability but at the same time creates conditions for the inter-helical interactions and the formation of the stable helical complex (Dong et al. 2012). So, Gly plays an intriguing role in peptide/protein structure formed in the membrane environment where they can act as tightly packing amino acids with a flexible main chain. It means that Gly is the best "aggregator" residue.

In some AMPs glycines are indeed considered as aggregators. Gly-xxx-Gly motives occur in several AMPs and they promote the self-aggregation or creation of the helical heterocomplexes. Studies of the amyloid-β peptide suggest that GxxxG glycine zipper motifs is responsible for dimerization and fibrillogenesis ( Kim 2009 ). It's known that GxxxG motif mediates the helix packing. At the same time this motif may be also critical for the formation and stability of β-sheet structure. Aβ peptide exists in a β-sheet or random coil configuration in the aquas environment, but converts to an α-helical structure upon membrane association (Fonte *et.al.* 2011). The same transitions have been shown for other glycine zipper containing peptides, plasticins (Bruston *et. al.* 2007; Carlier *et. al.* 2015) and bombinins (Zangger *et. al.* 2008; Petkov *et. al.* 2019 ).

Plasticins and bombinins are the members of Gly-rich peptides' family. The family of Gly rich antimicrobial peptides (GRAMP) united the peptides widely distinct by sequence and functionality. Gly-Leu rich peptides, such as XT7 peptide from Silurana tropicalis, acanthoscurrins 1 and 2, leptoglycin, etc are uncharged or weakly-charged linear peptides (Sousa *et. al.* 2009 ). Armadillidins with the presence of a sixfold repeated motif GGGF(H/N)(R/S) are highly cationic linear AMPs (Verdon *et. al.* 2016 ). While Gly-Cys rich peptides, such as ginsentides, are also uncharged or weakly-charged but have a highly compact, pseudocyclic structure stabilized by disulfide bond network (Tam *et al.* 2016 ). XT7 peptide, acanthoscurrins 1 and 2 and leptoglycin are active against Gram-negative bacteria and/or fungi( Sousa *et. al.* 2009 ). Armadillidins have equivalent antimicrobial activity against Gram-positive bacteria, Gram-negative bacteria and filamentous fungi, but not against yeasts ( Verdon *et. al.* 2016 ).

Consequently, it's difficult to suppose the common mode of action for GRAMPs. The only we can note is that, flexibility is a requirement to make linear peptide membrane-active. Glycines mainly provide the flexibility of linear peptides. In other cases, glycines provide bends in the polypeptide chains to create conditions for the formation of the disulfide-bonds network to stabilize the certain structural scaffold and to form flexible loops with special motives.

**5.5 Cysteines and AMPs. Intra-chain covalent bonds**

The majority of not-linear AMPs constitute of peptides stabilized their structure by the disulfide-bonds network. There are peptides that are shared a structural scaffold but can have different functionality due to differences in amino acid sequences. The conserved scaffold of a tertiary structure at the sequence variations allows the molecules to show different surface distributions of polar and apolar residues and a variation in the cationicity. Often the AMPs are grouped according to a common scaffold and/or to a defined number of cysteines. For instance, peptides displaying a pattern of six-cysteines are united into one group and the peptides with the pattern of eight-cysteines into another. The ubiquitous class of AMPs named defensins can be presented as the three groups of peptides displaying three different patterns of six -, eight-, and ten- cysteines (Sahl et al. 2005, Finkina and Ovchinnikova 2018). Correspondingly, scaffolds are stabilized by the network of three, four and five disulfide bonds. Classical hevein-like AMPs enriched in cysteine and glycine residues and stabilize structural scaffold by the network of 4 disulfide bonds; however, there are other sub-families with 6 or 10 cysteines, also (Slavokhotova et al. 2017).

Worth noting once more, that the same scaffold does not mean the same functionality. For instance, both hevein-like peptides (8C-HLP) and Ginsentides share a scaffold of eight cysteines that does not provide the same mode of activity of these peptides. 8C-HLPs have chitin-binding motives at inter-cysteine loop 3 and a conserved aromatic residue at loop 4, which are essential for chitin-binding and so, for antifungal activity (Kini et al. 2017). Ginsentides fail the ability to bind to chitin due to the absence of chitin-binding loops and so the specter of their targets is different from 8C-HLPs (Tam et al. 2018).

The structural scaffold of AMPs can be stabilized by other types of intrachain covalent bonds such as for instance thioether bonds in lantibiotics. Amino acid lanthionine serves the same function as the disulfide bridges in defensins. For the lantibiotic nisin the wedge model of action has been supposed (Driessen et al. 1995) similar to the toroidal pore model. But later has been shown that the nisins obviously use lipid II to bind specifically to the bacterial membrane, and the subsequent pore formation could proceed at much lower concentrations. For oyster defensins have been shown also, that they are specific inhibitors of a bacterial wall biosynthesis pathway rather than mere membrane-active agents; oyster defensin activity is a result of binding to lipid II (Schmitt et al. 2010). So it can be supposed that a stabilized by intrachain-bond rings ( loops) with the corresponding motives is a prerequisite for the binding to lipid II.

Two other bonds used for ring formation in AMPs are lactam and lactone bonds. A lactone ring is generated by cyclization of the C-terminal carboxylic acid with the side chain of serine or threonine, while cyclization between the C-terminal carboxylic acid or acidic side chains and the side chain of lysine or ornithine forms a lactam ring structures. AMPs with such intrachain bonds mainly synthesized by bacterial or fungal sources. Most antimicrobial cyclic peptides affect the integrity of the cell envelope (Lee and Kim 2015).

It's clear, that not linear peptides are more structured than linear ones due to stabilizing their tertiary scaffold by the intrachain-bonds network. Their flexibility is mainly linked with loops. So if loops possess corresponding motifs, the interactions of AMPs with each other or with other molecules (saccharides, proteins, lipids, etc) can be more specific. Consequently, we can say, that a mode of action of not linear AMPs mainly determined by the organization of the surfaces of the conservative scaffolds, for instance by the peculiarities of amino acid sequences of the loops.

## 6. Modes of AMP interaction with Biological membrane

In the coarse-grained approximation, an interaction of AMP with membrane can be finished by two kinds of results: permeabilization of membrane and/or penetration through it to reach an intracellular target, without membrane disruption.

**6.1 Permeabilization.**
The studies of the permeabilization of the membrane to understand the modes of action of AMP have taken place during several decades and relied on theoretical and experimental approaches. As the theoretical approaches MD simulations are mainly used (Sengupta *et al.* 2008; Wang *et al.* 2016; Ulmschneider 2017). The experiments were mainly performed on artificial membrane structures, such as large unilamellar vesicle (LUV) or giant unilamellar vesicle (GUV), using different biophysical methods, which mainly looked over the leaking capabilities of vesicles (Tamba *et.al*. 2010; Gregory *et al.* 2008; Oreopoulos *et al.* 2010; Dias *et al.* 2017). At the early stages of studies it was suggested, that most AMPs cause membrane permeabilization through one of three possible routes: carpet (Pouny *et al.* 1992 ), barrel-stave pore (He *et al.* 1996 ), or toroidal pore (Yang *et al.* 2001). Detail survey of these mechanisms has presented in several reviews (Hale and Hancock 2007; Lee, Hall and Aguilar 2016; Kumar, Kizhakkedathu, and Straus 2018). Despite a huge amount of work, a consensus understanding of AMP action is still lacking. Only for Alamethicin (Huang, Chen and Lee 2004; Qian et al., 2008) and lytic toxins (Matsuzaki, Yoneyama and Miyajima 1997; Sengupta et al. 2008) the formation of membrane-spanning pores is experimentally proved. These classical pore-forming peptides might be the exceptions. Consequently, the question is remaining open: What are the mechanisms behind all-or-none and graded leakage?

(Wimley and Hristova 2011). In many cases, leakage can be explained by transient-pores formation (Wang. et al. 2016). As the reason for these temporal not-structured pores the peptide-caused perturbation in the lipid arrangement with the consequence of phase transitions and appearing of the defects in the membrane has been considering. For instance for magainin 2 has been suggesting certain mechanism which implies high tension at the external monolayer of the membrane, caused by increasing its area by means of insertion of amphipathic peptides into it. An impact of the such tension on the internal monolayer can become the reason of the rupture of monolayer and sequential events forming pores stochastically. The radius of pores depend on magainin 2 concentration. (Tamba et.al. 2010). The experimental investigation of the mechanism of the release of the contents of phospholipid vesicles induced by another AMP, cecropin A, allowed to propose the model of pore-formation, an alternative to conventional (Gregory et al. 2008). This model doesn't require an oligomerization of peptides and suggests the disorganized structure of the pores which are not neatly lined by peptides. Even in the pore state of the vesicles, the fraction of peptides is retained on the membrane surface. So, according to the view of authors, cecropin A may cause membrane thinning and positive curvature strain, opening up pores that allow the complete release of contents. An unstable pore state of the vesicles is relaxed by lipid transport through the transient pores.

Consequently, there are views that other mechanisms that do not involve perforation of the membrane may exist. Results obtained from differential scanning calorimetry, nuclear magnetic resonance, and freeze-fracture microscopy studies (Epand et al. 2008; Jean-Francois et al. 2008) shows the clustering of the cationic antimicrobial peptide with certain anionic lipids can be the cause of the membrane crowding. These and other works have suggested that phase separation and/or domain formation may be an alternative mechanism of action for certain antimicrobial peptides. Atom force microscopy study of the small cationic AMP revealed that they induce the formation of cardiolipin-rich domains with a concomitant reduction in the ordering of the lipid acyl tails (Oreopoulos et al. 2010; Dias et al. 2017). This remodeling effect results in structural instabilities in the model membranes, suggesting phase separation as an alternative mechanism of antimicrobial peptide action. So, there is increasing evidence that the efficacy of some cationic antimicrobial agents is determined by their effects on membrane domains (Lohner 2009; Joanne et al. 2009). That is AMPs target specific membrane components and can induce specific restructuring of the membrane. The specific restructuring of the membrane can be the cause of the delocalization of peripheral membrane proteins impacting energy metabolism and/or cell-wall biosynthesis (Wenzel et al. 2014). Here it could be noted that a restructuring of the membrane is not only the basis of the new mechanisms that do not involve perforation but it also is a main theme in the conventional (old) models of AMP activity (Rakowska et al. 2013). It can be believed that preferential clustering of the cationic peptide with certain anionic lipids also provides an environment necessary to come to the conventional modes of membrane permeabilization (Pouny et al. 1992; He et al. 1996; Yang et al. 2001)

To uncover pore-formation mechanisms on atomic level the MD simulations of the systems of "peptide + bilipid" have been performed. The 21-residue AMP isolated from the skin of the green-eyed tree frog, in the bilayer of phosphatidylcholine lipids allowed to develop pore-forming mechanism distinct from some proposed models where AMP insertion was proposed via large surface aggregates (Sengupta et al. 2008). Maculatin induced membrane leakage is visualized experimentally (Ambroggio et al. 2005). MD simulation shows that Maculatin although initially reside in the state parallel to the surface, individual peptides subsequently adopt transmembrane (TM) orientation. An energetic barrier for Maculatin TM-insertion is overcome by cooperative actions, involving two peptides in a head-to-tail arrangement in combination with a water defect. At equilibrium, peptides are continually changing between marginally stable TM oligomeric assemblies and surface-bound states on both interfaces (Wang. *et al.* 2016). The author concluded that: pores form by consecutive addition of individual helices to a transmembrane helix or helix bundle; Maculatin forms an ensemble of structurally diverse temporarily functional low-oligomeric pores; These pores continuously form and dissociate in the membrane.

Today's knowledge of mechanisms of membrane permeabilization induced by AMP turn up a question: are there any common mechanisms at the interaction of the AMP with the membrane, followed leakage ? or there are many different mechanisms depending on physicochemical features of AMP, composition of membrane and their state. Anyway, the result of the interaction of AMP with membrane in the majority of cases is a leakage, which is taken place either through stable pores or through unstable, unstructured, transient pores. The formation of stable pores is proved experimentally for several AMPs only. Leakage induced by the majority of AMP is explained by transient defects (pores) in the membrane. We want to note, that in the current models, the transient pore formation is the cooperative process and often is accompanied by translocation of the part of peptides either to the internal monolayer of the membrane or into the cytoplasm (Shagaghi *et al.* 2017). Translocation into cytoplasm allows AMP to get intracellular, negatively charged targets ( such as DNA, RNA, proteins, mitochondrial membrane in the case of fungi and cancer cell, etc.)

### 6.2 Translocation

For many AMPs, pore formation capabilities depend on their concentration. For instance, PGLa at high concentration only can form pores, while at low concentration it adopts well-defined surface-bound S-state (Strandberg *et al*. 2009). Other examples are CPPs, which at high enough concentration perturb membranes and make it permeabilized (Palm, Netzereab and Hallbrink 2006) and Cecropin A, which at a low concentration may reach cytoplasmatic targets before the membrane permeabilization (Hong *et al.* 2003). Moreover for highly charged AMPs pore formation may not have to be considered at all to explain peptide translocation and membrane permeabilization. Peptide-lined pores are not needed at all to explain how a highly charged peptide can translocate across a lipid bilayer. Instead, simple cooperative effects involving 2–3 peptides can be used to explain translocation (Ulmschneider 2017 ). Indeed, it has been shown that some CPPs with large fractions of cationic residues are able to silently translocate membranes without causing too much leakage (He, Hristova and Wimley 2012; Ablan *et al.* 2016 ).

Therefore it is reasonable to suppose, that processes of translocation and permeabilization of the membrane by peptides have shared some common events and both processes are characterized by the cooperativity of peptides' actions on the membrane. Recently, it has been shown that some AMPs possess both capabilities, they can translocate through membrane and simultaneously form transient pores. As above mentioned for some such peptides result of interaction with membrane depends on the concentration of peptides (Strandberg *et al*. 2009; Palm, Netzereab and Hallbrink 2006; Hong *et al.* 2003). But in the case of Pur A both results, translocation, and permeabilization take place at the same peptide concentrations. It has been shown that peptide PuroA simultaneously passes through and creates pores in the membrane, although these two events are shifted in time (Shagaghi *et al.* 2017). It means that part of peptides pass through the plasma membrane before the pores will be formed. The results of several separate events (particular interactions) have to be determined by the relations between the kinetics of the events and the time necessary to gather enough actors for cooperative action.  In the case of PuroA, results are determined by the kinetics of the passing through the membrane and by the kinetics of the creation of transient aggregates for the pore-formation.

**6.3 Cooperation among AMPs.  Synergism**
Mode of action of particular AMP means sequential interactions of the peptide with the lipid bilayer and with other peptides. The results of interactions might be the aggregation of the peptide with others or just the creation of conditions for others to perform their mission easily (Herce and Garcia 2007). For instance, according to the mechanism of action supposed for Maculatin, the binding and insertion of the particular peptide into the membrane provide a lowering of the energy barrier of TM-insertion for other peptides (Wang. et al. 2016). In this context, it's clear that the mode and results of the interaction are peptide concentration-dependent (Strandberg et al. 2009; Palm, Netzereab and Hallbrink 2006). Consequently, in many models of AMP action, the interactions of the peptides with lipids and with each other provide conditions that are crucial to getting the final goal. Moreover, nature has provided the defense system of the organisms by the set of various AMPs, which act according to their own modes of actions and have certain potency, but at the same time can interact with each other and are strengthening the total potency (Marxer , Vollenweider and Schmid-Hempel 2016 ).  Latter fact presents an additional argument, that the ways to oppose the development of the resistance are well-optimized by the defense system based on the AMPs.

The strategy of the fighting against the development of the resistance used by nature becomes attractive to combat the resistance of microbes against previously active antibiotics. Current strategies to overcome the problem of the resistance to conventional antibiotics are not only the using of certain AMPs as a killer of the multi-drug resistant bacteria but also using them together with conventional antibiotics. Supposition about the success of combination therapy is supported by the facts that combination of drugs potentially eliminate resistant strains, delay the evolution of drug resistance, reduce the dosage of individual drugs, and hence, diminish side effects (Cokol et al 2011; Tamma, Cosgrove and Maragakis 2012; Worthington and Melander 2013). Recent studies report that success depends on the results of the combination, which can be synergistic or antagonistic (Yeh et al 2009; Chait , Craney and Kishony 2007). Synergistic drug pairs can efficiently kill bacteria but intensify selection of resistance, while antagonistic drug pairs showed the reverse trends. So the knowledge on the results of the interaction of combined drugs should be required.

Studies of the synergism of AMPs and conventional antibiotics to reveal a more effective combinations are intensively expanded (Feng  et.al. 2015; Wu et.al. 2017; Ruden  et.al. 2019 ;  Zhu  et.al. 2019; Vargas-Casanova  et.al. 2019 ). The results of the investigations are widely surveyed  (Hollmann et al. 2018 ). Here we will try to overview the knowledge concerning the synergisms between  AMPs only and  to present suppositions concerning the mechanisms of the synergistic actions.

An exploring of the pharmacodynamics of six different AMPs (cecropin A, LL 19-27 , melittin, pexiganan, indolicidin and apidaecin) by testing their individual and combined effects in vitro, allowed to conclude that the synergism is a common phenomenon in AMP interactions (Yu et al. 2016). Worth noting that three-AMP combinations are even more synergistic than two- AMP combinations (Yu et al. 2016).  Moreover, for Xenopus laevis (Bevins and Zasloff 1990) ,Tenebrio molitor (Johnston, Makarova and Rolff 2013) and bumble bee bombus terrestris (Marxer, Vollenweider and Schmid-Hempel 2016) have been shown that producing AMP cocktails is an efficient way to combat the bacterial invasion.

The phenomenon of synergism was looked over the AMPs of the Temporin family from the skin of Rana temporaria. Temporins A and B are active against gram-positive bacteria but not against gram-negative. While Temporin C is active against both groups of bacteria. It was supposed, that the resistance of gram-negative strains against Temporins A and B are linked with the capability of peptides to oligomerize in the presence of the outer membrane's LPS (Rosenfeld et al. 2006). Oligomerization hinders the moving towards the inner membrane. Interesting that Temporin C can prevent oligomerization of Temporins A and B promoted by the outer membrane. Therefore, when Temporin A and B were mixed with temporin C, a marked synergism observed (Rosenfeld et al. 2006).

PGLa and magainin 2 (MAG2) are the most well-studied AMPs from frog skin that show synergistic antimicrobial activity. The molecular mechanism of synergy between the last peptides was studied on the atomistic level by MD simulation (Pino-Angeles, Leveritt and Lazaridis 2016) and experimentally by different biophysical and biochemical methods (Matsuzaki et al. 1998; Strandberg et al. 2013; Zerweck et al 2016; Zerweck et al. 2017).

In the early work, Matsuzaki and coworkers have supposed that the synergistic action of magainin 2 and PGLa is a consequence of the formation of the heterodimeric complex (Matsuzaki et al. 1998). The complex was characterized by fast formation and moderate stability. It has been shown that the complex was not formed in the aqueous phase. Each peptide separately binds to the membrane and the complex is formed in the bilayer. The complex formation promotes the shift of the partitioning equilibrium of each component toward enhanced binding. So, it was supposed that the synergism is partially connected with the increased binding and formation of the heteromolecular complex (Matsuzaki et al. 1998). In another work (Zerweck et al 2016) has been suggested that the orientation of AMP relative to a lipid bilayer in the bound state predetermines a mode of action. Two states of AMP are considered: surface ("S-state") and transmembrane (inserted - "I-state"). For the given membrane, with particular curvature, the peptide state is concentration-dependent (Zerweck et al 2016). According to the heterodimer hypothesis mentioned above, it was proposed that PGLa and MAG2 formed a complex, where PGLa adopts transmembrane orientation, whiles MAG2 stays on the bilayer surface and stabilizes water-filled pore by interaction with PGLa (Strandberg et al. 2013).

To look for the amino acids necessary for synergy and to reveal details of molecular interactions the mutating variants of PGLa and MAG2 were explored (Zerweck et al. 2017). Zerweck and coworkers were using three different experimental approaches to explore the impact of amino acids at a certain position of the sequences on the different appearances of the synergy. Checker-board assays had allowed assessing the level of activity of the combined action of mutated variants of the PGLa and MAG2. By the solid-state 15N-NMR study was assessed the orientation of peptides relative to the lipid bilayer surface at the combined action. Leakage experiments were giving assessments of the sizes and stabilities of the defects in the lipid bilayer created by PGLa and MAG2 in cooperation. This multiapproach study allowed to reveal the GxxxG motif in the PGLa as a necessary to synergy. The motif is located between the polar and hydrophobic faces of the helical wheel. It's interesting, that GxxxG motifs are important for molecular contacts, but not just the motif, the exact position of the motives in the sequence is crucial also. In conclusion, the authors developed a molecular model of the functionally active PGLa-MAG2 complex, where Gly-Gly contact allows PGLa to form the membrane-inserted antiparallel dimer (Zerweck et al. 2017). The role of MAG2 is to promote the membrane-inserted state of PGLa and to stabilize the tetrameric heteropeptide complex. According to the model PGLa monomer is in contact with one MAG2 molecule at its C-terminus. Electrostatic interactions between anionic groups in MAG2 and cationic residues in PGLa are the basis of the last contact. Here worth noting that the study of the synergistic effects for other peptides derived by mutation allowed to conclude that though electrostatic interactions enhance synergy but are not necessary for the synergistic effect (Zerweck et al. 2017).

Along with the biophysical methods, molecular modeling and computer simulations are widely used to understand the behavior of the AMP in the membrane environment (Rzepiela *et al.* 2010; Woo and Wallqvist 2011; Han and Lee 2015; Pino-Angeles, Leveritt and Lazaridis 2016). Many efforts to describe the atomic level interactions taken place in the system composed of peptides and lipids have been performed. Worth noting, that the limited computational resources do not allow to perform a comprehensive description of complex systems composed of many molecules of lipids and peptides. Anyway, all-atom molecular dynamics simulations give valuable information about the details of the interactions of molecules. Recently the results of 5-9 μs all-atom molecular dynamics simulations of MAG2 and PGLa in DMPC or 3:1 DMPC/DMPG membranes have been reported (Pino-Angeles, Leveritt and Lazaridis 2016). Work was aimed to investigate a pore formation and stabilization process. Because of the limitation of the computational resources and unknown time scale for pore formation the simulations were starting from tetrameric helical bundles inserted in a transmembrane orientation. The rationale was that these starting conditions correspond to the local free energy minima of the pore-forming systems. Due to the fact that transmembrane state as starting for MAG2 is not convenient (Strandberg et al. 2013) and it's shown that a starting arrangement has an impact on the final structure of the pore (complex) (Pino-Angeles, Leveritt and Lazaridis 2016) we have to emphasize that last supposition can not be reasonable in the case of MAG2 homotetramer or heterotetramer of MAG2 -PGLa. Here worth noting that because of computational limitation, peptide to lipid molar ratio (P/L) in this simulation (Pino-Angeles, Leveritt and Lazaridis 2016) was equal or less than 1:30. It's known, that at such high peptide concentration PGLa has a propensity to adopt transmembrane, I –orientation (Strandberg et al. 2013). Consequently, the starting of simulation from inserted into membrane tetrameric helical bundles in the case of PGLa can be rational. But the same is not correct for MAG2 up to P/L 1:10 (Strandberg et al. 2013). Apparently, this is a reason why PGLa tetrameric assembly is stable nearly for the full MD trajectory and involves the four monomers in tilted T-states. While the simulation on the Mag2 tetramer shows that, two of the monomers adopt S state orientations on opposite leaflets during the first microsecond (Pino-Angeles, Leveritt and Lazaridis 2016). Nonetheless, the particular observations during MD simulation allow imaging of the mechanism of synergy between MAG2 and PGLa. For instance, interactions between residues S8 and E19 in MAG2 and K12 and K19 in PGLa, make the antiparallel heterodimer more stable than the homodimers (Pino-Angeles, Leveritt and Lazaridis 2016). So, MD simulation at the atomistic level shows that E19 of MAG2 indeed to be essential for maintaining the synergistic effect.

The synergistic effect of MAG2 on the insertion and aggregation of PGLa has been assessed in the coarse-grained molecular dynamics simulations (Han and Lee 2015). In these MD experiments, lipid bilayer constituted of dilauroylglycerophosphocholine (DLPC) lipids and the peptide/lipid molar ratio equaled about 0.023. The helical structure of peptides was fixed and not changed. At the initial stage of simulation, peptides were placed above the equilibrated bilayer surface with different heterodimeric orientations (parallel or antiparallel). Simulations showed that the synergistic effect from MAG2 promotes the tilting, insertion, and aggregation of PGLa. It has been indicated that PGLa aggregates only in the presence of MAG2, and not in the system without MAG2 (Han and Lee 2015). Authors supposed that MAG2 form parallel heterodimers with PGLa and induce the aggregation of heterodimers in the membrane. MAG2 tends to interact with the bilayer surface, while PGLa is tilted and inserted into the hydrophobic region of the bilayer to lead to the pore-formation (Han and Lee 2015).

## 7. Concluding remarks on the grouping of AMP

Many attempts have been performing to design compounds against drug resistance microbes relying on the features of AMP which are responsible for the mode of action. Due to difficulty to determine a features responsible for the particular mode of action, and absence of quantitative description of the mechanisms, the designing mainly performed experimentally and was relied on the known active peptides' sequences and on reasonable substitutions of the amino acids at the certain positions of parent peptide (Hilpert et al. 2005; Hai Nan et al. 2012; Haney et al. 2018). In these cases, design was coupled with structure-activity relationship studies. Another approach used is an *in silico* study relying on the machine–learning and the sets of data on experimentally explored peptides (Lee et al. 2017; Veltri, Kamath and Shehu 2018; Meher et al 2018; Thomas et al. 2009 ). The aim is to uncover the features of peptides which capable to distinguish one type of peptides from another and so to establish a predictive model. If we solve the task without taking into consideration the variety of the modes of action (which depends not only on peptides' properties but also on the type of envelope and environmental conditions), the solution will be the separation of two type of peptides, antimicrobial (AMP) and not-antimicrobial ( not - AMP), without specification of microbes against which peptide is active. To our knowledge, current predictive models mainly solve just such tasks, that is they used binary classification (Veltri, Kamath and Shehu 2018; Meher *et al*. 2018; Vishnepolsky and Pirtskhalava 2014; Thomas *et al.* 2009  ). The last task, in reality, is a gross-grained classification of peptides on antimicrobial and not-antimicrobial when a possibility of intra-AMP grouping is not considered.  For the target-oriented design more fine classification of peptides requires that take into consideration the capability of an intra-AMP grouping of peptides, for instance, according to results of interactions of AMP with envelope. The last interactions due to a variety of physicochemical features of peptides and their targets in the envelope come to the different results. The results of actions, that always start by binding of AMP with outer layer of envelope, maybe the following: a) Inhibition of the vital pathways in the outer layer of envelopes (for instance inhibition of the transferring of the cargo for peptidoglycan synthesis by Lipid II (Hasper *et al.* 2006 ) ); b) Permeabilization of the plasma membrane (for instance by disruption of membrane through pore-formation or carpet-formation (Pouny  *et al.* 1992; He  *et al.* 1996; Yang  *et al.* 2001  ), c) Perturbation of plasma membrane  structure without unrecoverable changes (For instance perturbation of the raft proteins and corresponding biological pathways through an interaction with lipid microdomains (Wenzel *et al.* 2014); d) translocation through plasma membrane ( without unrecoverable changes)  and  an inhibition of metabolic pathways in cytoplasm (Park, Kim and Kim 1998; Sharma *et al.* 2015).  Worth noting that the intra-AMP grouping according to the results of interaction with envelope is conditional. There are the peptides, whose interaction with the envelope gives several different results simultaneously. For instance, it has been shown that peptide *PuroA* simultaneously passes through the envelope and creates pores in the membrane (Shagaghi *et al.* 2017).

There were many different efforts to perform the intra-AMP grouping to raise the efficiency of the design. The intra-AMP grouping was tried to perform based on the data on sequences, structures,  target organisms, source organisms, etc. The designing was relying on many different encodings of sequence or structure.  Taking into consideration the high variability of AMPs in the particular organism and often their multifunctionality, the target-organism-based (Xiao *et al.* 2013 ) and source-organism-based (Chung *et al.* 2020) groupings could not be effective for the *de novo* designing.

The predictions based exclusively on sequences as a strings of letters and relying on the different machine-learning and so-called linguistic approaches are widely used for *de novo* design (Loose et al. 2006; Yoshida et al. 2018; Veltri, Kamath and Shehu 2018). Sequence-based binary classifications are details-ignoring approaches, because they are aiming to uncover common grammar in the set of AMPs which unites a groups of peptides with different modes of action and so with different grammars. Moreover, approaches based on sequence only are anyway accompanied by the loss of information valuable for the accurate description of the interactions with the membrane. If the whole 20 letter alphabet is used, the data on a similarity of amino acids ( between Lys and Arg for instance) is lost. In the case, when similar amino acids are grouped and used a simplified (short) alphabet, the loss of information (for instance on differences in physicochemical features of Lys and Arg ) takes place also.

The grouping of AMPs according to 3D structure distinguishes alpha-helical, beta-structural, alpha/beta, unstructured, etc. peptides. The majority of AMPs are linear and can adopt drastically different conformations depending on the environment. For instance, the beta-amyloid peptide is disordered in the water environment, creates alpha-helical oligomers in the membrane at low local concentration, but at high local concentration peptides are self-assembled into beta-structural complex (Landreh, Johansson and Jörnvall 2014). Another example: cationic plasticine, which has shown multiple conformational transitions, including destabilized helix states, beta-structures, and disordered states (El Amri and Nicolas 2008 ). Although the tertiary structures of intra-chain-bonded peptides have a more certain topology it's shown that similar tertiary scaffold does not mean similar functionality (Tam et al. 2018). So structural classification of AMPs is conditional and ambiguous.

In any case the mode of action of AMP and results of this action determined by the physicochemical properties (PCP) of peptide and composition of the cell envelope. Interactions between AMP and envelope can be described by simple physical forces, although these interactions might be finished by wide specters of different outcomes. Current data allow unambiguously distinguish AMPs according to physicochemical features. There are peptides with high value of total positive charge and peptides with zero or even negative charges; peptides with high hydrophobic moment and not - amphipathic; peptides with the propensity to transmembrane orientation and bound to membrane parallel. So, it's reasonable to think on the grouping of AMPs according to physicochemical features and to suppose the particular mode of action for each group of peptides. It has to be remembered, that it's right to do a grouping into the set of AMPs contains peptides active against particular strain (it means active against particular envelope). The effort to perform the clusterization of peptides according to physicochemical features has recently performed (Vishnepolsky et al. 2018; Vishnepolsky et al. 2019). Results of clusterization according to last works allow to suppose that physicochemical space of AMPs is discontinuous. At the same time due to the small size of the available sets of AMPs, the results can not be considered as reliable. So at the time being the question: are the physicochemical space of AMPs discrete ? is open. At some approximation, we can suppose that space is continuous. But to be sure about continuity more data on peptides to be active against particular strain is required. So up to the gathering enough data about peptides having potency against particular strains, it's rational to develop a predictive model relying on supposition about the discrecity of the physicochemical space. Consequently, at the time being, the model of prediction based on supposition about discrecity of the physicochemical space is optimal for the development of the method of the design with high performance.

So, aiming to develop a predictive model of high performance, it's reasonable to encode peptides on the base of physicochemical properties (PCP), that are responsible for the interaction with cell envelope and to try to perform the classification according to these properties. At the same time, during classification, we have to remember, that particular peptide with certain physicochemical properties (PCP) can interact differently with the different envelopes. Even the mode of action against the same envelope, in the same conditions, can be multimodal, because the kinetics of gathering of the peptide at the different sites of the envelope can be different. Consequently, due to the differences in local concentration of the peptide on different site of the membrane, the results of action of the same peptide on the different sites of membrane can be different: from the penetration into cytoplasm (at low local concentrations) to the aggregation (self- or through lipids) and destruction of membrane (pore- or carpet- formation at high local concentrations).

Here the question arises: what is an optimal set of physicochemical descriptors that should be used? Authors have used many different descriptors to classify AMP, but we can say that all of them are the derivatives of charge and hydrophobicity. For instance, such descriptor as the propensity of AMP to aggregation depends on the composition and distribution of charged and hydrophobic residues along the chain, the same we can say on a hydrophobic moment, a propensity to disordered structure, isoelectric point and other descriptors used. We have to emphasize that the proper attention has not taken such marked physical interaction as the cation-π interactions. A comprehensive understanding of the role of last interaction at the functioning of AMP is crucial for instance to answer on the questions: What is a cause of necessity of abundance of aromatic residues (including Trp) in the AMPs? and why does nature preferably uses Lys and not Arg when designing AMPs? While it's shown that Arg can enhance either the membrane permeability or translocating capability of AMP (Cutrona *et al.* 2015 ). As above mentioned, Lys and Arg are differently interacting with cell membranes and with aromatic residues. Guanidinium group of Arg is a major cause of this distinction. Special binding of the Arg to phosphate groups of lipids or rings of aromatic residues is explained by the specific structure of the guanidinium group (Robison *et al.* 2016). The imagine of Arg as an important residue to rise an activity, gives further motivation to increase Arg content in the *de novo* designed AMPs (Cutrona *et al.* 2015 ). Due to the last tendency, a portion of Arg in synthetic peptides is remarkably high than in ribosomal peptides (see "statistics" page of DBAASP https://dbaasp.org/statistics).
In this work, we suppose that an abundance of Lys in ribosomal AMPs connected with the necessity to rise a selectivity of AMPs to the prokaryotic cell membrane. But this is the supposition only and feature study of the role of Arg and Lys in the interactions with lipid's phosphate or headgroups and aromatic residues of amino acids is necessary to understand more comprehensively the modes of action and to develop the models of *in silico* design of AMP with high performance. To our best knowledge, there are not used a decoding system at the classification of AMP that is relied on descriptors that reflect the peculiarity of Arg an Lys in the light of cation-π interactions. So it might be noted that an effective classifier is not yet developed and the process of looking for the optimal set of descriptors to classify AMPs is continuing.